\documentclass{article}
\usepackage{url}
\DeclareUnicodeCharacter{2032}{\ensuremath{^{\prime}}}
\usepackage{jheppub} 
\usepackage{lineno}
\usepackage{braket}
\usepackage{url}
\usepackage{hyperref}
\usepackage{doi}

\title{Spread and circuit complexity as a measure of particle content and phase space fluctuations}

\author[1,2]{Satyaki Chowdhury,}
\affiliation[1]{Institute of Theoretical Physics, Jagiellonian University, Lojasiewicza 11, 30-348 Cracow, Poland}
\emailAdd{satyaki.chowdhury@doctoral.uj.edu.pl}
\affiliation[2]{Doctoral School of Exact and Natural Sciences, Jagiellonian University, Lojasiewicza 11, 30-348 Cracow, Poland}

\abstract{In this work, we investigate the relation between different notions of quantum complexity, namely, circuit and spread complexity and physically meaningful quantities such as the particle content of the quantum state and the variances of position and momentum operators. Using a harmonic oscillator with time-dependent mass and frequency as a toy model, we show that both circuit and spread complexity at any instant is determined by the mean number of quanta and its rate of change. Furthermore, both complexity and its growth are directly linked to the variances of the position and momentum operators, providing a clear physical interpretation of complexity in terms of the state's excitation and phase-space fluctuation. Although the analysis is carried out for a single time-dependent oscillator, the results have direct relevance for quantum field theory in curved backgrounds, where individual field modes effectively behave as time-dependent oscillators. This offers new insights into how quantum complexity encodes particle production and phase space fluctuations in non-holographic systems. Finally, we establish a precise and potentially universal relation between spread and circuit complexity for the time evolved state suggesting deeper connections between different complexity measures in the context of field theories on curved backgrounds.
}

\begin{document}
\maketitle
\flushbottom

\section{Introduction}
\label{sec:intro}

One of the concepts of quantum information theory that has taken the central stage in the recent past is complexity, which is used to measure the difficulty of performing a certain task. Surprisingly, it is found to play a significant role in understanding black hole interiors. It was suggested that complexity is the manifestation of the linear growth of the black hole interiors with time, probed by extremal surfaces \cite{Susskind:2014rva}. Such statements have been tested in various scenarios in the context of AdS/CFT \cite{Susskind:2014rva,Brown:2015bva,Stanford:2014jda,Carmi:2016wjl,Belin:2021bga} and have already provided supporting evidence.

Various definitions of quantum complexity have been proposed that attempt to accurately capture this notion of difficulty \cite{Aaronson:2016vto,Jefferson:2017sdb,Caputa:2017yrh,Bhattacharyya:2018wym,Caputa:2018kdj,Bhattacharyya:2018bbv,Ali:2018fcz,Bhattacharyya:2021cwf,Ali:2018aon,Osborne:2012mft,Chapman:2017rqy,Chagnet:2021uvi,Chen:2020nlj,Cotler:2017jue,Brandao:2019sgy,Haferkamp:2021uxo,Bouland:2019pvu,Munson:2024usy,Balasubramanian:2019wgd,Chowdhury:2023iwg}. Two of the most popular and widely used ones that will also be of interest to us are that of \textit{Nielsen complexity} \cite{Nielsen_2006,https://doi.org/10.48550/arxiv.quant-ph/0502070,https://doi.org/10.48550/arxiv.quant-ph/0701004} and \textit{Krylov complexity} \cite{Parker:2018yvk}. Nielsen et. al. developed a geometrical approach to determine complexity of $n$-qubit unitary transformation. Within this geometric framework, the complexity of the desired unitary transformation was identified with the length of the shortest geodesic between the two points denoting the identity and the target unitary operation, respectively, in the space of unitary operators. The shortest geodesic gives the optimal circuit for the transformation. This geometric approach was utilized to define the notion of complexity of quantum states \cite{Jefferson:2017sdb,Chapman:2017rqy} and was thereafter applied in various quantum mechanical, field theoretic and cosmological settings \cite{Bhattacharyya:2020art,Bhattacharyya:2020kgu,Bhattacharyya:2020rpy,Bhattacharyya:2024duw,Bhattacharyya:2024rzz,Bhattacharyya:2025cxv,Bhattacharyya:2018bbv,Bhattacharyya:2019kvj,Bhattacharyya:2020iic,Bhattacharyya:2022rhm,Bhattacharyya:2021fii,Erdmenger:2020sup,Bhattacharyya:2022ren,Bhattacharyya:2023sjr}. A generalization of the approach to arbitrary unitary operators was provided in \cite{Chowdhury:2023iwg} and was utilized in \cite{Chowdhury:2024ntx,Bhattacharyya:2024rzz}.

Krylov complexity on the other hand measures how an initial local operator at some initial time $t=0$ evolves to some complex operator generated by the Hamiltonian $H$. The growth of operator measures how an initially localized operator spreads over the entire system. This notion of complexity has been extensively used in the recent past to study thermalization and probe the integrable to chaotic transition in quantum many body systems \cite{Parker:2018yvk,Barbon:2019wsy,Rabinovici:2020ryf,Nandy:2023brt,Dymarsky:2021bjq,Avdoshkin:2022xuw,Camargo:2022rnt}. It was explored in a variety of spin chain models including the SYK \cite{Rabinovici:2022beu,Rabinovici:2021qqt,Bhattacharya:2023xjx,Bhattacharjee:2022ave} and the Bose-Hubbard models \cite{Bhattacharyya:2023dhp}. The idea of Krylov complexity is based on a recursive technique called the Lanczos algorithm. Using this algorithm, an orthogonal basis (called the Krylov basis) and a set of Lanczos coefficients are obtained. One then studies the growth of the operator in this basis. The idea of complexity in Krylov space was extended to quantum states in \cite{Balasubramanian:2022tpr} and is termed as \textit{Spread complexity}. It quantifies the spread of an initial quantum state under an Hamiltonian in the Krylov basis. This has given new directions in the detection of quantum phase transitions in chaotic systems \cite{Balasubramanian:2022tpr,Bhattacharjee:2022qjw,Caputa:2022eye}. The investigation of spread complexity was extended to open quantum systems and non-hermitian models, which requires a generalization of the Lanczos algorithm \cite{Bhattacharya:2022gbz, Bhattacharya:2023xjx, Bhattacharya:2023yec,Bhattacharya:2023zqt,Bhattacharyya:2023grv,Carolan:2024wov, Liu:2022god, Chakrabarti:2025hsb, Bhattacharyya:2025lsc,Baggioli:2025knt}. Interested readers are referred to the reviews \cite{Nandy:2024evd,Baiguera:2025dkc,Chapman:2021jbh} and references therein for more details.  

Although there has been a flurry of works exploring the role of complexity in various quantum mechanical and field theoretic settings, it's relation to physically interpretable quantities remains unexplored. Some advancement in this direction was made within the AdS/ CFT correspondence. A conjecture stating that the rate of growth of complexity is proportional to the radial momentum of massive particles in AdS spacetime \cite{Susskind:2018tei,Susskind:2019ddc,Brown:2018kvn,Susskind:2020gnl,Magan:2018nmu,Barbon:2019tuq,Barbon:2020uux}, i.e.
\begin{align}
\label{eq1}
    \frac{dC(t)}{dt} \propto P,
\end{align}
where $C(t)$ refers to a complexity measure capturing the size of Heisenberg operators $O(t)$ which grows following the Heisenberg equation of motion i.e:
\begin{align}
    O(t)= e^{iHt}O(0)e^{-iHt}.
\end{align}
The dual description of the effective size of the Heisenberg operators is given by falling particles with momentum $P$ in AdS spacetime. The conjecture written in Eq. \ref{eq1} was tested using the notion of operator growth in Krylov basis and Krylov complexity. To be precise, the rate of growth of spread complexity of locally excited 2D CFT states were computed and matched with the proper radial momentum of the dual massive particles in AdS${}_3$ \cite{Caputa:2024sux}. The case for massless particles was investigated in \cite{Fan:2024iop} and both the cases were revisited in \cite{He:2024pox}.

With the motivation of testing relations like Eq. \ref{eq1} and establishing relations between quantum complexity and physically interpretable quantities in non-holographic theories, we investigate the model of an oscillator with time-dependent mass and frequency. We study both Nielsen's complexity and spread complexity to study the system and relate both complexity measures with the time-dependent particle content of the state. Moreover, we show that these measures can be expressed in terms of the expectation values of the position and momentum operators. Finally, we explore a possible relation between the two notions of complexity in the context of quantum field theory on curved backgrounds.

The organization of the paper is as follows. In Section \ref{sec2}, we discuss the general formalism of an oscillator with time-dependent parameters (mass and frequency). We discuss the general form of the time evolved state of the system and how the notion of time-dependent particle content arises in this context. In section \ref{sec3}, we present the analysis of Nielsen's complexity by applying the covariance matrix approach. We show how Nielsen's complexity can be related to the so- called excitation parameter, which in turns allows us to relate complexity with the time-dependent particle content of the state. In section \ref{sec4}, we discuss the notion of spread complexity for the system in consideration. We show that when the Hamiltonian is expressible in terms of the elements of a Lie algebra, it is possible to utilize the so-called Lie algebra decoupling theorem and construct the Krylov basis. However, owing to the time dependence in the Hamiltonian, the Lanczos coefficients will be time-dependent. 
The Lie algebra decoupling theorem  and some useful derivations are presented in appendices \ref{appendixA}, \ref{appendixB}, and \ref{appC} respectively.

\section{General formalism of a time-dependent oscillator}
\label{sec2}
The quantum harmonic oscillator with time- dependent parameters is ubiquitous in the study of a quantum field in a non-trivial time-dependent background and hence has been widely investigated in numerous contexts.
The typical approach in quantum field theory in a non-trivial time-dependent background is to take a semi-classical approach where a quantized degree of freedom $q$ (e.g a scalar field) interacts with a classical degree of freedom, $C$ (e.g. a cosmological background). The quantum theory of $q$ is determined by the configuration of $C$. If the configuration is non-trivial, the theory of $q$ is based on a time-dependent Hamiltonian. 

A harmonic oscillator with time-dependent mass and frequency is described by the following Hamiltonian:
\begin{align}
\label{Hamiltonian}
    H= \frac{p^2}{2 m(t)}+ \frac{1}{2}m(t)\omega(t)^2 q^2,
\end{align}

The equation of motion of this oscillator can be written as 
\begin{align}
\label{equationofmotion}
    \frac{d}{dt}(m(t)\dot{q})+ m(t)\omega(t)^2 q=0.
\end{align}

The Hamiltonian Eq. \ref{Hamiltonian} can describe a particular Fourier mode of a quantum field in a time-dependent classical background. For example, a quantum field in FRW spacetime characterized by a scale factor $a(t)$ or an electric field expressed in a time-dependent gauge with vector potential $A(t)$.

The evolution of the quantized oscillator, satisfying the time-dependent Schrodinger equation, is described by the wavefunction $\psi(q,t)$, as:
\begin{align}
\label{timedependentschreq}
    i\frac{\partial \psi(q,t)}{\partial t}= -\frac{1}{2m(t)}\frac{\partial^2\psi(q,t)}{\partial q^2}+\frac{1}{2}m(t)\omega(t)^2 q^2 \psi(q,t).
\end{align}
In this context, it is beneficial for us to work in the Schrodinger picture as we will be dealing with quantum states. One can also work in the Heisenberg picture where the evolution is described by the time-dependent creation and the annihilation operators. 
For a time-dependent oscillator, the concept of a unique vacuum or ground state exist only when the time-dependent parameters describing the oscillator goes to a constant value asymptotically. However, this is a special scenario and is mostly not true, for example, the frequency of a field mode in the Friedmann universe. In general, there is no concept of unique vacuum. However, there exist a class of solutions to the time-dependent oscillator which are \textit{form-invariant} \cite{Mahajan:2007qc,Mahajan:2007qg,Wigner:1932eb,Albrecht:1992kf,Matacz:1992mk,Padmanabhan:1986hda,Polarski:1995jg,Sriramkumar:2004pj} in the sense that the $q$ dependence is the same at all times. The most general state having this property is an exponential of a quadratic function of $q$ and is given by:
\begin{align}
\label{forminvariant}
    \psi(q,t)= N(t) \exp(-R(t) q^2).
\end{align}

It can be easily seen that this form of the state provides a solution to \ref{timedependentschreq}, which can be interpreted as the ground state of the oscillator at some initial time $t_0$. Eqn. \ref{forminvariant} is popularly known as the squeezed quantum state and is extensively studied in the context of quantum optics \cite{Caves:1985zz,Schumaker:1986tlu} and cosmology \cite{Grishchuk:1989ss,Grishchuk:1990bj,Grishchuk:1994sj,Albrecht:1992kf}.

Substituting \ref{forminvariant} into the time-dependent Schrodinger equation \ref{timedependentschreq}, we get the following equations for $N$ and $R$:
\begin{align}
\label{eqnNandR}
    i \frac{\dot{N}}{N}= \frac{R}{m}, ~~~~ i\dot{R}= \frac{2R^2}{m}-\frac{m \omega^2}{2}.
\end{align}

The normalization function $N(t)$ satisfies the condition:
\begin{align}
\label{normalization}
    |N|^2= \sqrt{\frac{R+R^*}{\pi}}.
\end{align}
The equation for $R$ (first order, nonlinear) can be identified as the generalized Riccati type and hence can be transformed into second order linear equation by introducing a new parameter $\mu$ as:
\begin{align}
    R= -i \frac{m}{2}\frac{\dot{\mu}}{\mu}.
\end{align}
 Substituting in \ref{eqnNandR}, it can be verified that $\mu$ satisfies the following differential equation:
 \begin{align}
     \Ddot{\mu}+\frac{\dot{m}}{m}\dot{\mu}+\omega^2 \mu=0,
 \end{align}
 which is identical to the classical equation of motion satisfied by the oscillator variable $q$.

 Before presenting the exact solution, it is instructive to obtain the adiabatic limit, in which the time-dependent functions $m(t)$ and $\omega(t)$ are slowly varying functions of time, such that the terms $\frac{\dot{m}}{m} << 1$ and $\frac{\dot{\omega}}{\omega}<<1$. Therefore, in this limit, the equation for $\mu$ becomes:
 \begin{align}
     \Ddot{\mu}+\omega(t)^2 \mu \approx 0.
 \end{align}
 The solutions to the above equation can be approximated by the WKB ansatz as:
 \begin{align}
     \mu(t) \approx \frac{1}{\sqrt{\omega(t)}}\exp\bigg(i \int_0^t \omega(t') dt'\bigg).
 \end{align}
 Thus, in the adiabatic limit, we have:
 \begin{align}
     R(t) \approx \frac{m(t)\omega(t)}{2}.
 \end{align}
 Substituting the solution of $R(t)$, the time-dependent normalization factor, $N(t)$,  can be written as:
 \begin{align}
     N(t)= C \exp\bigg(-\frac{i}{2}\int_{t_0}^t \omega(t') dt'\bigg).
 \end{align}
These equations determine the evolution of the state in the adiabatic approximation. 
Eq. \ref{forminvariant}, show that the entire information of the quantum state is encoded in the time dependence of the function $R(t)$. Since we have $R(t) \approx \frac{m(t)\omega(t)}{2}$ in the adiabatic limit, the exact solution can be determined by measuring how $R(t)$ deviates from the adiabatic approximation. To measure the deviation, it is convenient to introduce a complex function $z(t)$ which is related to $R(t)$ by the relation:
\begin{align}
\label{Rformdeviation}
    R(t)=\frac{m(t) \omega(t)}{2}\bigg(\frac{1-z}{1+z}\bigg) .
\end{align}

This is motivated by the form of $R(t) \approx \frac{m(t)\omega(t)}{2}$ in the adiabatic approximation. The function $z(t)$ clearly measures the deviation of $R(t)$ from the adiabatic value and therefore can be termed the\textit{ excitation parameter}. In terms of $z$, the time evolved state can be written as: 
\begin{align}
\label{wavefunctionz}
    \psi(q,t)= N(t)\exp\bigg(-\frac{m(t)\omega(t)}{2}\bigg(\frac{1-z}{1+z}\bigg)q^2\bigg).
\end{align}
From the differential equation satisfied by $R(t)$, one can obtain the equation obeyed by $z$ which satisfies the following differential equation:
\begin{align}
    \dot{z}+2i \omega z+ \frac{1}{2}\bigg(\frac{\dot{\omega}}{\omega}+\frac{\dot{m}}{m}\bigg)(z^2-1)=0.
\end{align}


The variable $z(t)$ completely determines the state of the system and allows one to define a set of variables with reasonable physical interpretation, built out of the wavefunction. These quantities are crucial from the point of view of understanding the physical content of the quantum state. One such quantity is the particle content of the state. 

When the parameters of the oscillator are time-independent, the particle content of the state can be defined in the usual sense. However, for an oscillator with time-dependent parameters, one cannot define stationary states, and hence the usual notion of particle content do not exist. A reasonable and physically motivated way of quantifying the time-dependent content of the state would be to compare it with the instantaneous energy eigenstates at a given moment. The wavefunction of the oscillator that started off in the instantaneous ground state at some instant $t_0$, will at a later time $t$ be in the superposition of the instantaneous eigenstates defined at that moment. This non-zero probability of being in the $n$-th energy eigenstate at $t$ can be interpreted as the excitation of quanta.    

The instantaneous eigenstates at time $t$ can be defined as:
\begin{align}
    \phi_n(q,t)= \bigg(\frac{m(t)\omega(t)}{\pi}\bigg)^{1/4}\frac{1}{\sqrt{2^n n!}}H_n(\sqrt{m(t)\omega(t)}q) \exp\bigg\{-\frac{m(t)\omega(t)}{2}q^2\bigg\},
\end{align}
which correspond to the eigenfunctions of the Hamiltonian \ref{Hamiltonian}, with the parameters fixed at $t$. However, these states do not correctly describe adiabatic evolution. Under adiabatic evolution, a state prepared in the instantaneous ground state at some initial time $t_0$ will evolve to coincide with the instantaneous ground state at every subsequent moment $t$. To accurately incorporate this feature, the appropriate instantaneous eigenstates must contain an additional phase factor along with $\phi_n(q,t)$. Therefore, the appropriate eigenstate defined at a time $t$ is given by:
\begin{align}
\nonumber
    \psi_n(q,t)&= \exp\bigg(-i\int_{t_0}^t E_n(t')dt'\bigg) \phi_n(q,t) \\
    &= \bigg(\frac{m(t)\omega(t)}{\pi}\bigg)^{1/4}\frac{1}{\sqrt{2^n n!}}H_n(\sqrt{m(t)\omega(t)}q) \exp\bigg\{-\frac{m(t)\omega(t)}{2}q^2-i \int_{t_0}^t \bigg(n+\frac{1}{2}\bigg)\omega(t')dt'\bigg\}.
\end{align}
 
The wavefunction at any time $t$, will be in a superposition of instantaneous eigenstates defined at that moment and can be written as:
\begin{align}
    \psi(q,t)= \sum_{n=0}^{\infty}C_n(t)  \psi_n(q,t).
\end{align}

The time-dependent coefficients $C_n(t)$ can be calculated as follows:
\begin{align}
    C_n(t) &= \int_{-\infty}^{\infty} \psi_n^*(q,t)\psi(q,t) dq \\
    &= N \bigg(\frac{m\omega}{\pi}\bigg)^{1/4}\frac{1}{\sqrt{2^n n!}} \exp\bigg(-i\int_{t_0}^t (n+\frac{1}{2})\omega dt\bigg)\int_{-\infty}^{\infty} dq H_n(\sqrt{m\omega}q)e^{-\bigg(R+\frac{m\omega}{2}\bigg)q^2}.
\end{align}

The quantities $N, m, \omega$ and $R$ in the above equations are time-dependent quantities but we have avoided writing it explicitly for notational simplicity.

\begin{align}
\label{hermiteintegral}
    C_n(t)= f(t) \int_{-\infty}^{\infty}dq H_n(\sqrt{m\omega}q) e^{-a q^2},
\end{align}
where the quantities $f(t)$ and $a$ in the above equation are given by:
\begin{align}
    f(t)&= N \bigg(\frac{m\omega}{\pi}\bigg)^{1/4}\frac{1}{\sqrt{2^n n!}} \exp\bigg(-i\int_{t_0}^t (n+\frac{1}{2})\omega dt\bigg), \\
    a &= R+\frac{m\omega}{2}.
\end{align}

The integral in \ref{hermiteintegral} can be carried out using the generating function of the Hermite polynomials \cite{enwiki:1317601709}, which is given by:
\begin{align}
    G(x,t)= e^{-t^2+2xt}= \sum_{n=0}^{\infty}\frac{t^n}{n!}H_n(x).
\end{align}
Introducing $\sqrt{m\omega}q=x$, we get:
Multiplying both sides of the above equation, with $e^{-a q^2}$ and integrating we get:
\begin{align}
    \int_{-\infty}^{\infty}e^{-t^2+2 t\sqrt{m\omega}q}e^{-aq^2} dq = \sum_{n=0}^{\infty} \frac{t^n}{n!}\int_{-\infty}^{\infty}H_n(\sqrt{m\omega}q)e^{-aq^2}dq.
\end{align}

Using the standard Gaussian integrals, we get:
\begin{align}
\nonumber
    \sum_{n=0}^{\infty} \frac{t^n}{n!}\int_{-\infty}^{\infty}H_n(\sqrt{m\omega}q)e^{-aq^2}dq &= \sqrt{\frac{\pi}{a}}\exp\bigg\{t^2\bigg(\frac{m\omega}{a}-1\bigg)\bigg\} \\
    &= \sqrt{\frac{\pi}{a}}\sum_{k=0}^{\infty}\frac{1}{k!}\bigg(t^{2k}\bigg(\frac{m\omega}{a}-1\bigg)^k\bigg).
\end{align}

Equating coefficients of equal power of $t$ on both sides, we see that the integral contributes only when $n=2k$. Therefore, we have 
\begin{align}
    I_{2k}= \int_{-\infty}^{\infty}H_{2k}(\sqrt{m\omega}q)e^{-aq^2}dq= (2k)! \sqrt{\frac{\pi}{a}}\frac{(\frac{m\omega}{a}-1)^k}{k!}.
\end{align}

The non-zero time-dependent coefficients are therefore given by:
\begin{align}
    C_{2k}= N \bigg(\frac{m\omega}{\pi}\bigg)^{1/4}\frac{1}{\sqrt{2^{2k}(2k)!}}\exp\bigg(-i\int_{t_0}^t \bigg(2k+\frac{1}{2}\bigg)\omega dt'\bigg) I_{2k}.
\end{align}

The probability for the oscillator to be in the eigenstate $\psi_{2k}$ at time $t$ is given by:
\begin{align}
    P_{2k}(t)= |C_{2k}|^2= M \frac{(2k)!|z|^{2k}}{(k!)^2 2^{2k}},
\end{align}
where $M$ in the above equation refers to:
\begin{align}
    M= \frac{|N|^2\sqrt{\pi m \omega}}{|R+\frac{m\omega}{2}|}.
\end{align}

Writing $|N|^2$ and $R$ in terms of $z$, $M$ can be expressed as:
\begin{align}
    M= \sqrt{1-|z|^2}.
\end{align}

The mean number of quanta (here only the even quanta contributes) in the state at time $t$ is given by:
\begin{align}
\label{mean quanta}
    \langle n(t) \rangle = \sum_{k=0}^{\infty}2k P_{2k} = \sum_{k=0}^{\infty}2k \sqrt{1-|z|^2}\frac{(2k)!|z|^{2k}} {(k!)^2 2^{2k}}= \frac{|z|^2}{1-|z|^2}.
\end{align}
This quantity $\langle n \rangle$ can be interpreted as the ``particle content" of the quantum state at the instant $t$. Although it is not a unique definition, it is certainly a reasonable one. For a time-dependent Hamiltonian, the mean value of the energy at time $t$ can be computed by taking the expectation value of the Hamiltonian $E(t)= \bra{\psi}H \ket{\psi}$. Using, Eqns.\ref{forminvariant}, \ref{normalization} and \ref{Rformdeviation}, it can be shown that $E(t)$ is given by:
\begin{align}
    E(t)= \omega(t) \bigg(\frac{1+|z|^2}{1-|z|^2}\bigg).
\end{align}

Using the definition of $\langle n \rangle$ from Eq. \ref{mean quanta}, it can be shown that:
\begin{align}
    E(t)= \omega(t)\bigg(\langle n \rangle +\frac{1}{2}\bigg).
\end{align}

Although the mean particle number just depends on the magnitude of $z$, it does not contain the entire information of the quantum state. In terms of $z$, the quantum state can be written as:
\begin{align}
    \psi(q,t)= N(t) \exp\bigg\{-\frac{m(t)\omega(t)}{2}\bigg(\frac{1-z}{1+z}\bigg)q^2\bigg\}.
\end{align}
Since $z$ is a complex quantity, it can be written as:
\begin{align}
    \psi(q,t)= N(t) \exp\bigg\{-\frac{m(t)\omega(t)}{2}\bigg(\frac{1-|z|e^{i\theta}}{1+|z|e^{i\theta}}\bigg)q^2\bigg\}.
\end{align}

Thus, a complete description of the quantum state requires not only the magnitude of $z$ but also the phase factor $\theta$. 

The dynamical equations can also be expressed in terms of $\langle n \rangle$ and $\theta$ as follows:
\begin{align}
\label{ndot}
    \langle \dot{n} \rangle &= \bigg(\frac{\dot{\omega}}{\omega}+\frac{\dot{m}}{m}\bigg)\sqrt{\langle n \rangle(\langle n \rangle+1)}\cos(\theta), \\
    \label{thetadot}
    \dot{\theta} &= -2 \omega -\frac{1}{2}\bigg(\frac{\dot{\omega}}{\omega}+\frac{\dot{m}}{m}\bigg)\frac{2\langle n \rangle+1}{\sqrt{\langle n \rangle(\langle n \rangle+1)}}\sin(\theta).
\end{align}

It will be useful to calculate the time-dependent expectation value of the position and momentum operator, which will be of particular use to us. Due to the Gaussian nature of the wavefunction, it is obvious that $\langle q(t)\rangle= \langle p(t) \rangle =0$, therefore the significant ones to us are $\langle q(t)^2\rangle$ and $\langle p(t)^2 \rangle$, which can be calculated as:
\begin{align}
\label{positionavg}
    \langle q(t)^2 \rangle &= \frac{1}{2 m(t)\omega(t)}\bigg\{ \frac{1+|z|^2+2 Re(z)}{1-|z|^2}\bigg\},\\
    \label{momentumavg}
    \langle p(t)^2 \rangle &= \frac{1}{2}m(t)\omega(t)\bigg\{ \frac{1+|z|^2-2 Re(z)}{1-|z|^2}\bigg\}.
\end{align}

In terms of $\langle n(t) \langle$ and $\langle \dot{n}(t)\rangle$, the expectation values can be re-expressed as:
\begin{align}
\label{eqnqsqavg}
    \langle q(t)^2 \rangle &= \frac{1}{2m(t)\omega(t)}\bigg\{(2\langle n(t) \rangle+1)+2\frac{\langle \dot{n}(t) \rangle}{\frac{\dot{\omega}(t)}{\omega(t)}+\frac{\dot{m}(t)}{m(t)}}\bigg\}, \\
    \label{eqnpsqavg}
    \langle p(t)^2 \rangle &=\frac{1}{2}m(t)\omega(t)\bigg\{(2\langle n(t) \rangle+1)-2\frac{\langle \dot{n}(t) \rangle}{\frac{\dot{\omega}(t)}{\omega(t)}+\frac{\dot{m}(t)}{m(t)}}\bigg\}.
\end{align}

With all the physically meaningful quantities at hand, we now proceed to the computation of both circuit complexity and spread complexity and investigate their relations these quantities. 

\section{Circuit complexity of the time evolved state}
\label{sec3}
The notion of circuit complexity in quantum field theory essentially quantifies the effort of preparing a target quantum state $\ket{\Psi_T}$ starting from a certain reference state $\ket{\Psi_R}$. This means constructing the shortest quantum circuit that performs the transformation
\begin{align}
    \ket{\Psi_T}= U \ket{\Psi_R},
\end{align}
where the unitary is constructed as a sequence of elementary unitary gates:
\begin{align}
    U= \prod_{n=1}^{N}g_n .
\end{align}
$N$ represents the total number of gates present in the circuit. The total number $N$ of gates present in the optimal construction is referred to as complexity or better \textit{gate complexity}. Generally, arbitrarily many different quantum circuits, i.e. different sequences of the $g_i$'s, are possible that yields the same $\ket{\Psi_T}$. So, the primary challenge lies in identifying the optimal circuit amongst the infinite possible constructions. Nielsen et. al. in a series of papers \cite{https://doi.org/10.48550/arxiv.quant-ph/0502070,https://doi.org/10.48550/arxiv.quant-ph/0701004,Nielsen_2006} introduced a geometric way to identify the optimal circuit. In this geometric approach, the problem of finding the optimal circuit is related to finding minimal length curves in the space of unitary operations. The length of the shortest geodesic between the identity and the desired unitary operation gives a measure of its complexity. 

Since, we will deal with quantum states and not unitary operators, it is necessary to identify the difference between the notion of ``state" complexity and ``unitary" complexity. In ``unitary" operator complexity, the unitary operator is the standalone object and one is interested in the complexity of that particular operator. However, in state complexity, the object of interest is the target quantum state, and its complexity depends on the choice of the initial reference state. The choice of quantum gates required to construct the quantum circuit depends on the choice of reference state. 

Let us briefly highlight the key features of the notion of state complexity. Given an initial reference state and a set of elementary gates, the task is to construct the most efficient quantum circuit that starts at the reference state and ends at the desired target state. Therefore,
\begin{align}
    \ket{\Psi_T}= U(s)\ket{\Psi_R},
\end{align}
where $U(s)$ is the unitary operator representing the quantum circuit and is written as:
\begin{align}
    U(s)= \mathcal{P}\exp\bigg(i \int_0^s ds' H_{\rm eff}(s')\bigg).
\end{align}
$s$ parametrizes a path in the space of unitaries. Therefore, the problem is to find the effective $s$ dependent Hamiltonian that synthesizes the desired $U$. The effective Hamiltonian can be expanded in a certain basis $O_I$ as:
\begin{align}
    H_{\rm eff}(s)= V^I(s)O_I.
\end{align}
The basis $O_I$'s represent the generators of the elementary gates and $V^I(s)$ are the so-called control functions that specifies the tangent vector to a trajectory in the space of unitaries. The idea is to then define a \textit{cost} for the various possible paths as:
\begin{align}
\label{eqncost}
    D(U(s))= \int_0^1 ds F(U(s),\dot{U}(s))~.
\end{align}
Minimization of this cost functional leads to identification of the optimal set of $V^I(s)$, which in turn gives us the optimal circuit. There are various possible choices for the cost function \cite{Nielsen_2006,https://doi.org/10.48550/arxiv.quant-ph/0502070,https://doi.org/10.48550/arxiv.quant-ph/0701004,Jefferson:2017sdb,Guo:2018kzl,Hackl:2018ptj,Khan:2018rzm}. However, in this paper, we consider $F_q= \sqrt{\sum_I q_I (V^I)^2}$. Using this cost functional, Eq. \ref{eqncost} can be suitable expressed as
\begin{align}
    D(U(s))= \int_0^1 ds \sqrt{G_{IJ}V^IV^J}~ .
\end{align}
The above equation shows that the optimal trajectory corresponds to the geodesic in the corresponding geometry.

In this paper, we will focus on the notion of ``state" complexity. Recently, the notion of operator complexity has been extensively used to investigate the complexity of time evolution operators in various contexts \cite{Balasubramanian:2019wgd,Balasubramanian:2021mxo,Chowdhury:2023iwg,Chowdhury:2024ntx,Bhattacharyya:2024rzz,Haque:2021hyw,Haque:2024ldr}. Since, the reference and the target states under consideration are Gaussian, we will adopt the covariance matrix approach to do the complexity analysis. However, there are other approaches to studying the complexity of quantum states, such as the Fubini-Study method \cite{Chapman:2017rqy}. Interested readers are referred to \cite{Ali:2018fcz} for a comparative analysis of the various approaches to the complexity of Gaussian states.    


A pure Gaussian state can be parameterized by its wavefunction as 
\begin{align}
    \psi(q)= \bigg(\frac{a}{\pi}\bigg)^{1/4} \exp\bigg[-\frac{1}{2}(a+ib)q^2\bigg],
\end{align}
with $a$, $b$ being real valued. All the information of a Gaussian state is contained in the quadratic combination of the position and momentum coordinates. Grouping the position and the momentum coordinate as $\xi^m=(x,p)$, we get
\begin{align}
    \bra{\psi}\xi^m\xi^n\ket{\psi} =\frac{1}{2}(G^{mn}+i \Omega^{mn}),
\end{align}
where $G^{mn}$ is symmetric and $\Omega^{mn}$ is anitisymmetric, which can be completely fixed by the canonical commutation relation as
\begin{align}
    \Omega= \begin{pmatrix}
        0 && 1 \\
        -1 && 0
    \end{pmatrix}.
\end{align}
The symmetric covariance matrix $G$, with entries $G^{mn}$ has entries given by the expectation values as follows: 
\begin{align}
\label{covariancematrix}
    G= \begin{pmatrix}
        2\langle q^2 \rangle && \langle qp+pq \rangle \\
        \langle qp+pq \rangle && 2\langle p^2 \rangle
    \end{pmatrix}
    = \begin{pmatrix}
        \frac{1}{a} && -\frac{b}{a} \\
        -\frac{b}{a} && \frac{a^2+b^2}{a}
    \end{pmatrix}.
\end{align}
Eq. \ref{covariancematrix}, shows that the entire information of the Gaussian wave functions is encoded in the covariance matrix. 

Our reference state is the ground state of the oscillator at $t=0$, i.e, 
\begin{align}
    \psi_R= N(t=0) \exp\bigg(-\frac{m_0 \omega}{2}q^2\bigg).
\end{align}

Simialrly, our target state is the time evolved state at any time $t$, which can be written as:
\begin{align}
    \psi_T= \psi(q,t)= N(t)\exp(-\frac{\omega_t(t)}{2}q^2),
\end{align}
where $\omega_t= m(t) \omega(t)\frac{1-z}{1+z}$.
The covariance matrices for the reference and target states $\psi_R$, $\psi_T$ takes the following form:
\begin{align}
    G_R= \begin{pmatrix}
        \frac{1}{m_0\omega} && 0 \\
        0 && m_0 \omega
    \end{pmatrix}, ~~~~
    G_T= \begin{pmatrix}
        \frac{1}{Re(\omega_t(t))} && -\frac{Im(\omega_t(t))}{Re(\omega_t(t))} \\
        -\frac{Im(\omega_t(t))}{Re(\omega_t(t))} && \frac{|\omega_t(t)|^2}{Re(\omega_t(t))}
    \end{pmatrix}.
\end{align}

Introducing a symmetric matrix $S$, the reference and target covariance matrices can be simplified so that the reference covariance matrix becomes identity. The effect of the matrix $S$, is as follows:
\begin{align}
    SG_RS^T= \mathbb{I}, ~~~~ \Tilde{G_T}= S G_T.S^T .
\end{align}
In this case, the matrix $S$ can be found to be:
\begin{align}
    S= \begin{pmatrix}
        \sqrt{m_0 \omega} && 0 \\
        0 && \frac{1}{\sqrt{m_0\omega}}
    \end{pmatrix}.
\end{align}

With the above transformation, the target covariance matrix becomes:
\begin{align}
    \Tilde{G_T}= \begin{pmatrix}
        \frac{m_0 \omega}{Re(\omega_t(t))} && -\frac{Im(\omega_t(t))}{Re(\omega_t(t))} \\
        -\frac{Im(\omega_t(t))}{Re(\omega_t(t))} && \frac{|\omega_t(t)|^2}{m_0 \omega Re(\omega_t(t))} 
    \end{pmatrix}.
\end{align}

The generators that are suitable for our purpose are as follows: \cite{Ali:2019zcj}:
\begin{align}
    O_1= \frac{i}{2}(qp+pq), ~~~ O_2= \frac{i}{2}q^2, ~~~ O_3= \frac{i}{2}p^2
\end{align}

These will serve as the generators of the elementary gates that will be used to construct the circuit and satisfy the $SL(2,R)$ lie algebra with the following commutation relations:
\begin{align}
    [O_1,O_2]= 2 O_2, ~~ [O_1,O_3]= -2O_3, ~~ [O_2,O_3]= O_1 .
\end{align}

Using the set of elementary gates written above, the complexity of $\ket{\psi_T}$ with respect to $\ket{\psi_R}$ can be calculated as \cite{Ali:2019zcj}:
\begin{align}
\label{circuitcomplexity}
    C= \frac{1}{2}{\rm arcosh}\bigg(\frac{1}{2}\frac{| \omega_t(t)| ^2+m_0^2 \omega^2}{m_0 \omega Re(\omega_t(t))}\bigg).
\end{align}


Substituing $\omega_t= m(t)\omega(t)\frac{1-z}{1+z}$, Eq. \ref{circuitcomplexity} can be re-expressed as:
\begin{align}
\nonumber
    C &= \frac{1}{2}{\rm arcosh}\bigg(\frac{1}{2}\frac{m(t)^2\omega(t)^2 \frac{1+|z|^2- 2 Re(z)}{1+|z|^2+2 Re(z)}+m_0^2 \omega^2}{m_0 \omega m(t)\omega(t) \frac{1-|z|^2}{1+|z|^2+2 Re(z)}}\bigg) \\ \label{complexityintermsofz},
    &= \frac{1}{2}{\rm arcosh}\bigg(\frac{1}{2}\frac{m(t)^2\omega(t)^2(1+|z|^2-2 Re(z))+m_0^2\omega^2(1+|z|^2+2 Re(z))}{m_0\omega m(t)\omega(t)(1-|z|^2)}\bigg).
\end{align}

Without loss of generality, we take the initial values of $m(t)$ and $\omega(t)$ i.e $m_0$ and $\omega_0$ to be 1. From equations \ref{positionavg} and \ref{momentumavg}, Eq. \ref{complexityintermsofz} can be written as:
\begin{align}
\label{circuitcomplexitypandq}
    C(t)= \frac{1}{2}{\rm arcosh}\bigg( \langle q(t)^2 \rangle+ \langle p(t)^2 \rangle\bigg).
\end{align}

It is also useful to express the complexity in terms of mean number of particles at any instant. From Eqn. \ref{eqnqsqavg} and \ref{eqnpsqavg}, it can be shown that the complexity can also be re-expressed as:
\begin{align}
\label{circuitcomplexityparticle}
    C(t)= \frac{1}{2}{\rm arcosh}\bigg(\mathcal{A}(t) (2\langle n(t) \rangle+1)+ \mathcal{D}(t)\langle \dot{n}(t) \rangle\bigg).
\end{align}


where the time-dependent functions $\mathcal{A}(t)$ and $\mathcal{D}(t)$ are given by:
\begin{align}
    \mathcal{A}(t) = \frac{1}{2}\frac{m(t)^2\omega(t)^2+1}{ m(t)\omega(t)}, ~~~~~~~\mathcal{D}(t)=\frac{\frac{1-m(t)^2\omega(t)^2}{m(t)\omega(t)}}{\frac{\dot{\omega}(t)}{\omega(t)}+\frac{\dot{m}(t)}{m(t)}}.
\end{align}

At $t=0$, the expectation values can be calculated as:
\begin{align}
    \langle q(t=0)^2 \rangle= \frac{1}{2}, ~~~~ \langle p(t=0)^2 \rangle= \frac{1}{2},
\end{align}
which gives:
\begin{align}
    C(t=0)=0.
\end{align}

This is expected since at $t=0$, the reference and the target states are identical and hence the complexity should be 0.


The relation between circuit complexity and the mean number of quanta in the state given in Eq. \ref{circuitcomplexityparticle} can equivalently be expressed in terms of the mean energy and its rate of change as:
\begin{align}
    C(t)=\frac{1}{2}{\rm arcosh}\bigg(\mathcal{F}(t)E(t)+ \mathcal{G}(t)\dot{E}(t)\bigg),
\end{align}

where $\mathcal{F}(t)= \frac{2 \mathcal{A}(t)-\mathcal{D}(t)\dot{\omega}(t)}{\omega(t)}$, and $\mathcal{G}(t)= \frac{\mathcal{D}(t)}{\omega}$.

The rate of change of complexity can be calculated as:
\begin{align}
\label{complexityratechange}
    \frac{dC(t)}{dt}= \frac{1}{2}\frac{\dot{Q}(t)+\dot{P}(t)}{\sqrt{(Q(t)+P(t))^2-1}},
\end{align}
where $P(t)$ and $Q(t)$ in the above expression refer to $\langle p(t)^2 \rangle$ and $\langle q(t)^2 \rangle$, respectively. 

Eq. \ref{complexityratechange} presents a direct relation between the rate of change of circuit complexity and the time variation of the second moments of the canonical operators. Therefore, within the context of quantum field theory on nonstationary backgrounds, this relation demonstrates that the growth of circuit complexity is governed by fluctuations in position and momentum.






\section{Spread complexity of the time evolved state}
\label{sec4}
Spread complexity measures the rate at which an initial state spreads through the Hilbert space of a quantum dynamical system. To quantify this measure, a cost function is introduced, which tracks the spread of the initial state over all the possible basis. The true measure of complexity then lies in minimizing the cost function over all possible choice of basis \cite{Balasubramanian:2022tpr}. In \cite{Balasubramanian:2022tpr}, it was proved that a unique orthonormal basis minimizes this cost function throughout a finite time interval for continuous evolution and for all times in the case of discrete evolution. This is known as the \textit{Krylov basis}. This unique basis can be obtained by an implementation of the Lanczos recursion algorithm \cite{Lanczos:1950zz}, which was conceived as a method for tridiagonalizing Hermitian matrices.   

The Lanczos algorithm is a powerful iterative procedure used to construct an orthonormal basis for the Krylov subspace associated with a Hermitian operator. Starting from a reference vector $\ket{v_0}$, and a Hermitian operator $H$, 
the Krylov subspace of order $n$ generated by $H$ and $\ket{v_0}$ is defined as:
\begin{align}
    \mathcal{K}_n(H,\ket{v_0})= {\rm span}\big\{\ket{v_0},H\ket{v_0},H^2\ket{v_0},...., H^{n-1}\ket{v_0}\big\}.
\end{align}

The primary goal of the Lanczos algorithm is to generate an orthonormal basis $\{\ket{v}_n\}$ for this subspace such that the action of the operator $H$ is tridiagonal in this basis, which is achieved by a three term recurrence relation:
\begin{align}
\label{iteration}
    \ket{\Tilde{v}_{n+1}}= b_{n+1}\ket{v_{n+1}}= H\ket{v_n}-a_n\ket{v_n}-b_n\ket{v_{n-1}},
\end{align}
where $a_n= \langle v_n|H|v_n \rangle$ and $b_{n+1}=\sqrt{\langle \Tilde{v}_{n}|\Tilde{v}_n \rangle}$.

Rearranging Eq. \ref{iteration} we get:
\begin{align}
\label{iterationgeneral}
    H\ket{v_n}= b_n\ket{v_{n-1}}+a_n\ket{v_n}+ b_{n+1}\ket{v_{n+1}}.
\end{align}

The time evolving state in the minimizing Krylov basis can be written as:
\begin{align}
\label{PsiinKrylovbasis}
    \ket{\Psi(t)}= \sum_{n=0}^{K-1} \psi_{n}(t)\ket{v_n},
\end{align}
where $K$ is the dimension of the span of the time-evolving state and $\ket{v_0}=\ket{\Psi(0)}$. In quantum mechanics, for a time independent Hamiltonian, the time evolution operator $U(t)=e^{-iHt}$ can be used to generate the Krylov basis. However, for the time-dependent case, things might become non-trivial owing to the presence of the time ordering in the evolution operator. When the Hamiltonian is time-dependent, one may consider constructing a time-dependent Krylov basis at each instant of time $t$. Such an approach to construct the time-dependent Krylov basis was done in \cite{Takahashi:2024hex} by utilizing the so called $(t,t')$ formalism.

However, for time-dependent Hamiltonians possessing an underlying Lie algebraic structure, the analysis of complexity in the Krylov basis becomes significantly simplified. To be more specific, when the time-dependent Hamiltonian admits a representation in terms of the generators of a Lie algebra, one can bypass the Lanczos recursion algorithm for constructing the Krylov basis. Within the group representations of such algebras, the Hamiltonian takes a tridiagonal form, allowing a direct identification of the representation basis with the Krylov basis. However, the transition weights or the Lanczos coefficient will be time-dependent. 


Let us illustrate the above argument with the illustrative example of the following time-dependent Hamiltonian which is expressible in terms of the generators of $\mathfrak{su}(2)$ Lie algebra:
\begin{align}
    H(t)= \alpha(t) (J_{+}+J_{-})+\gamma(t)J_0+ \delta(t)I,
\end{align}
where $J_{\pm}$ and $J_0$ are the generators of $\mathfrak{su}(2)$, satisfying the following commutation relation 
\begin{align}
    [J_0,J_{\pm}]= \pm J_{\pm}, ~~~~~ [J_{+},J_{-}]=2J_0.
\end{align}
The spin states $\{j,-j+n\}$ with $n=0,1,....,2j$ automatically follows the Lanczos algorithm and tridiagonalize the Hamiltonian $H(t)$, which can be understood from the action of the generators $J_{\pm}$ and $J_0$ on the state $\{\ket{j,-j+n}\}$. 
\begin{align}
    J_0 \ket{j.-j+n} &=(-j+n)\ket{j,-j+n}, \\
    J_{+}\ket{j.-j+n} &= \sqrt{(n+1)(2j-n)}\ket{j,-j+n+1},\\
    J_{-}\ket{j.-j+n} &= \sqrt{n(2j-n+1)}\ket{j,-j+n-1}.
\end{align}

Therefore, the action of the Hamiltonian on the state $\ket{j,-j+n}$, gives:
\begin{align}
\nonumber
    H(t)\ket{j,-j+n} &= \alpha(t)\sqrt{(n+1)(2j-n)}\ket{j,-j+n+1} \\ \nonumber &~~~ + \alpha(t)\sqrt{n(2j-n+1)}\ket{j,-j+n-1}  + \gamma(t)(-j+n)\ket{j,-j+n} \\ \nonumber & ~~~~ + \delta(t)\ket{j,-j+n}, \\ \nonumber
    &= \alpha(t)\sqrt{n(2j-n+1)}\ket{j,-j+n-1}+ \big(\gamma(t)(-j+n) \\ & ~~~~~~  +\delta(t)\big)\ket{j,-j+n} + \alpha(t)\sqrt{(n+1)(2j-n)}\ket{j,-j+n+1}.
\end{align}

Comparing, the above equation with Eq. \ref{iterationgeneral}, the Lanczos coefficients can be written as:
\begin{align}
    a_n(t)&= \gamma(t)(n-j)+\delta(t), \\
    b_n(t) &= \alpha(t)\sqrt{n(2j-n+1)}.
\end{align}

The above illustrative example show that for time-dependent Hamiltonian linear in the generators of a Lie algebra, an exact time-independent Krylov basis can be constructed from the representation theoretic considerations with time-dependent Lanczos coefficients. This approach is extremely beneficial for extending the Krylov complexity considerations for time-dependent Hamiltonians.

To apply the above formalism in our case, it is essential to show that the Hamiltonian \ref{Hamiltonian} is expressible in terms of the generators of a Lie algebra. For this purpose, we introduce the creation and annihilation operators at the initial time $t=0$:
\begin{align}
    \hat{a} &= \sqrt{\frac{m_0\omega_0}{2}}\hat{q}+ \frac{i}{\sqrt{2m_0\omega_0}}\hat{p}, \\
    \hat{a}^{\dagger} &= \sqrt{\frac{m_0\omega_0}{2}}\hat{q}- \frac{i}{\sqrt{2m_0\omega_0}}\hat{p},
\end{align}
where $m_0$ and $\omega_0$ are $m(t=0)$ and $\omega(t=0)$. 
Using this set of creation and annihilation operators defined at $t=0$, we can write down eqn \ref{Hamiltonian} as:
\begin{align}
    \hat{H}(t)= \bigg(\frac{m(t)\omega(t)^2}{4 m_0\omega_0}-\frac{m_0\omega_0}{4m(t)}\bigg)(\hat{a}^2+\hat{a}^{\dagger 2})+ \bigg(\frac{m(t)\omega(t)^2}{4 m_0\omega_0}+\frac{m_0\omega_0}{4m(t)}\bigg)(\hat{a}\hat{a}^{\dagger}+\hat{a}^{\dagger}\hat{a}).
\end{align}
Defining: 
\begin{align}
    K_{-}= \frac{\hat{a}^2}{2}, ~~~ K_{+}= \frac{\hat{a}^{\dagger 2}}{2}, ~~~ K_3= \frac{\hat{a}\hat{a}^{\dagger}+\hat{a}^{\dagger}\hat{a}}{2}.
\end{align}
It is straightforward to check that the above operators satisfy the $\mathfrak{su}(1,1)$ Lie algebra as follows:
\begin{align}
    [K_{+},K_{-}]= -2 K_3, ~~~ [K_3,K_{\pm}]= \pm K_{\pm} .
\end{align}
In terms of these $K_i$'s, the Hamiltonian can be rewritten as:
\begin{align}
\nonumber
    \hat{H} 
     \label{HamiltonianLiealgebra}
    &= \bigg(\frac{m(t)^2\omega(t)^2-m_0^2\omega_0^2}{2m_0\omega_0 m(t)}\bigg) (K_{+}+K_{-}) + \bigg(\frac{m(t)^2\omega(t)^2+m_0^2\omega_0^2}{2m_0\omega_0 m(t)}\bigg)K_3.
\end{align}
Defining $\lambda(t)$, $\lambda_0(t)$ as:
\begin{align}
    \lambda(t) = \bigg(\frac{m(t)^2\omega(t)^2-m_0^2\omega_0^2}{2m_0\omega_0m(t)}\bigg),~~~~
    \lambda_0(t) = \bigg(\frac{m(t)^2\omega(t)^2+m_0^2\omega_0^2}{2m_0\omega_0m(t)}\bigg),
\end{align}
the Hamiltonian \ref{HamiltonianLiealgebra} can be re-expressed as:
\begin{align}
\label{Hamiltoniangenerators}
    H(t)= \lambda(t)(K_{+}+K_{-})+ \lambda_0(t)K_3.
\end{align}

In the realization of the time-dependent oscillator in terms of the $\mathfrak{su}(1,1)$ generators, the Hilbert space naturally decomposes into two subspaces corresponding to the even and odd Fock states. Each subspace forms a lowest weight representation of $\mathfrak{su}(1,1)$, known as the positive discrete series, labeled by the Bargman index $h$ which takes values $1/4$ and $3/4$ for the even and odd sector respectively. The ground state of the oscillator coincides with the lowest weight state $\ket{h,0}$ of the even representation satisfying $K_{-}\ket{h,0}=0$. The higher states of the representation are generated by the repeated action of $K_{+}$. The action of the generators $K_{+}$, $K_{-}$ and $K_{0}$ on $\{\ket{h,n}\}$ are as follows:
\begin{align}
    K_{+}\ket{h,n} &=\sqrt{(n+1)(2h+n)}\ket{h,n+1},\\
    K_{-}\ket{h,n} &=\sqrt{n(2h+n-1)}\ket{h,n-1},\\
    K_{3}\ket{h,n} &=(h+n)\ket{h,n}.
\end{align}
Within this representation basis, the Hamiltonian $\ref{Hamiltoniangenerators}$ becomes diagonal as follows:
\begin{align}
\nonumber
    H(t)\ket{h,n} &= \lambda(t)\sqrt{(n+1)(2h+n)}\ket{h,n+1}+ \lambda(t)\sqrt{n(2h+n-1)}\ket{h,n-1} \\ & ~~~~~~~~ + \lambda_0(t)(h+n)\ket{h,n}.
\end{align}
Consequently, the representation basis of the positive discrete series coincides with the Krylov basis for this system, with the following time-dependent Lanczos coefficients: 
\begin{align}
    a_n(t)= \lambda_0(t)(h+n), ~~~  b_n= \lambda(t)\sqrt{n(2h+n-1)}.
\end{align}

Eq. \ref{HamiltonianLiealgebra}, shows that the Hamiltonian of an oscillator with time-dependent mass and frequency can be written in terms of the generators of the $\mathfrak{su}(1,1)$ Lie algebra. Since, the Hamiltonian is time-dependent, the time evolution operator contains a non-trivial time ordering which takes care of the non-commutativity of the Hamiltonian at different times. However, due to the presence of this time ordering, investigating the dynamics becomes difficult and one usually resorts to a perturbative approach via the Dyson series. However, an efficient way of investigating the dynamics of a quantum system (characterized by a time-dependent Hamiltonian) that lives in an infinite-dimensional Hilbert space but could be generated by a finite Lie algebra is provided by the so-called \textit{Lie algebra decoupling theorem}, which we briefly review in appendix \ref{appendixA}. The Lie algebra decoupling theorem allows us to express the time evolution operator as product of exponential operators with time-dependent coefficients. Whenever, analytical solutions of the time-dependent coefficients exist, the exact dynamics of the quantum system is known.


For the Hamiltonian written in \ref{HamiltonianLiealgebra},
\begin{align}
    H(t)= \lambda(t) (K_{+}+ K_{-})+ \lambda_0(t) K_3 +\delta  \mathbb{I},
\end{align}
employing the Lie algebra decoupling theorem, the evolution operator can be written in a decomposed form as:
\begin{align}
\label{decoupledevolution}
    U(t)= e^{\Gamma_{+}(t)K_{+}}e^{\Gamma_{3}(t)K_{3}}e^{\Gamma_{-}(t)K_{-}},
\end{align}
where $\Gamma_{+}, \Gamma_{3}$ and $\Gamma_{-}$ are functions of $\lambda(t)$ and $\lambda_{0}(t)$ and can be determined by substituting Eq. \ref{decoupledevolution} into the Schrodinger equation \ref{schrodingereqnevolution} and using the Baker-Campbell Hausdorff formula.



The action of the evolution operator on the lowest weight state $\ket{h,0}$, can be written as:
\begin{align}
    \ket{\Psi(t)} &= e^{\Gamma_{+}(t)K_{+}}e^{\Gamma_{3}(t)K_{3}}e^{\Gamma_{-}(t)K_{-}}\ket{h,0}, \\
    &= e^{\Gamma_3(t)h} e^{\Gamma_{+}(t)K_{+}}\ket{h,0},\\
    &= e^{\Gamma_3(t)h} \sum_{n=0}^{\infty}\frac{(\Gamma_+(t))^n K_{+}^n}{n!}\ket{h,0}, \\
    & = e^{\Gamma_3(t)h}\sum_{n=0}^{\infty} \frac{\Gamma_{+}(t)^n}{n!}\sqrt{\frac{\Gamma(2h+n)}{\Gamma(2h)}}\sqrt{n!}\ket{h,n},\\
    &= e^{\Gamma_3(t)h}\sum_{n=0}^{\infty}\Gamma_{+}(t)^n\sqrt{\frac{\Gamma(2h+n)}{\Gamma(2h)n!}}\ket{h,n}.
\end{align}

For SU(1,1), $\Gamma_3(t)$ is related to $\Gamma_{+}$ as:
\begin{align}
    {\rm Re(\Gamma_3)}= \ln(1-|\Gamma_{+}|^2)
\end{align}
The derivation of this relation is relegated to Appendix \ref{appendixB}.

Rewriting the above equation, we get:
\begin{align}
\label{Psiinkrylovourmodel}
     \ket{\Psi(t)}=\underbrace{e^{Re(\Gamma_3(t)+i Im(\Gamma_3(t))h}\sum_{n=0}^{\infty}\Gamma_{+}(t)^n\sqrt{\frac{\Gamma(2h+n)}{\Gamma(2h)n!}}}_{\psi_n(t)} \underbrace{\ket{h,n}}_{\text{Krylov basis}}.
\end{align}

The complexity in the Krylov basis is given by:
\begin{align}
\label{complexityspread}
    C_S= \sum_{n=0}^\infty n |\psi_n(t)|^2,
\end{align}
where $\psi_n(t)$ for our case can be recognized by comparing \ref{Psiinkrylovourmodel} to \ref{PsiinKrylovbasis}. Therefore, Eq. \ref{complexityspread} can be written as:
\begin{align}
    C_S= (1-|\Gamma_3(t)|^2)^{2h} \sum_{n=0}^{\infty}n (\Gamma_{+}(t))^{2n}\frac{\Gamma(2h+n)}{\Gamma(2h)n!}.
\end{align}

Simplifying the above expression, we get:
\begin{align}
\label{KrylovintermsofGamma}
    C_S(t)= 2h \frac{|\Gamma_{+}(t)|^2}{1-|\Gamma_{+}(t)|^2}
\end{align}

Therefore, the time evolution of spread complexity is completely determined by the explicit form of the time-dependent function $\Gamma_{+}(t)$. As already pointed out earlier, the explicit forms of $\Gamma_{\pm}$ and $\Gamma_{3}$ can be calculated by substituting the disentangled form of the evolution operator written in \ref{decoupledevolution} in the Schrodinger equation. However, their explicit forms will not be important in our case. Instead, what we want is to relate the spread complexity with $\langle p(t)^2 \rangle$ and $\langle q(t)^2 \rangle$ and hence, it is useful to find the relation between $\Gamma_3(t)$ with the excitation parameter $z$. 

Using this evolution operator, we can evolve our initial state which was the ground state of the oscillator at the initial time:
\begin{align}
    \ket{\psi(t)} &= U(t)\ket{0}= e^{\Gamma_{+}(t)K_{+}}e^{\Gamma_{3}(t)K_{3}}e^{\Gamma_{-}(t)K_{-}} \ket{0}, \\
    &= e^{\Gamma_{+}(t)K_{+}}e^{\Gamma_{3}(t)K_{3}}\ket{0}, \\
    &= e^{\Gamma_{+}(t)K_{+}}e^{\frac{\Gamma_3(t)}{4}}\ket{0},\\
    &= e^{\frac{\Gamma_3(t)}{4}}\sum_{n=0}^{\infty}\frac{(\Gamma_{+}(t))^n}{2^n n!}\sqrt{2n!}\ket{2n}.
\end{align}
The state $\ket{0}$ and $\ket{2n}$ in the above derivation denotes the ground state and the $2n$-th energy eigenstate at the initial time $t_0$. 
It is now useful to express $\ket{\psi(t)}$ in the position basis. The purpose of this is to relate the parameter $R(t)$ introduced in \ref{forminvariant} to the time-dependent quantities $\Gamma(+)$, $\Gamma(-)$ and $\Gamma(3)$. 
Therefore, we have:
\begin{align}
\label{psiqt}
    \psi(q,t)= e^{\frac{\Gamma_3(t)}{4}}\sum_{n=0}^{\infty}\frac{\Gamma_+(t)^n}{2^n n!}\sqrt{2n!}~\psi_{2n}(q).
\end{align}
Substituting $\psi_{2n}(q)$:
\begin{align}
    \psi_{2n}(q)= \frac{1}{\sqrt{2^{2n}(2n)!}}\bigg(\frac{m_0\omega_0}{\pi}\bigg)^{1/4}e^{-\frac{m_0\omega_0q^2}{2}}H_{2n}(\sqrt{m_0\omega_0}q),
\end{align}
in Eq. \ref{psiqt}, we get:
\begin{align}
\label{wavefunctionGamma}
    \psi(q,t)= e^{\frac{\Gamma_3(t)}{4}}\sum_{n=0}^{\infty}\frac{\Gamma_+(t)^n}{2^n n!}\sqrt{2n!}\frac{1}{\sqrt{2^{2n}(2n)!}}\bigg(\frac{m_0\omega_0}{\pi}\bigg)^{1/4}e^{-\frac{m_0\omega_0q^2}{2}}H_{2n}(\sqrt{m_0\omega_0}q).
\end{align}
The above expression can be suitably re-expressed as:
\begin{align}
\label{psiqtwavefunction}
    \psi(q,t)= e^{\frac{\Gamma_3(t)}{4}}\bigg(\frac{m_0\omega_0}{\pi}\bigg)^{1/4}\frac{1}{\sqrt{1+\Gamma_{+}(t)}}\exp\bigg\{-\frac{m_0\omega_0}{2}\bigg(\frac{1-\Gamma_{+}(t)}{1+\Gamma_{+}(t)}\bigg)q^2\bigg\}.
\end{align}
The derivation of the above form of the wavefunction is shown in Appendix \ref{appC}.
On equating Eq. \ref{psiqtwavefunction} to Eq. \ref{wavefunctionz}, we get the following relation between the time-dependent quantity $\Gamma_{+}(t)$ and the excitation parameter $z$:
\begin{align}
\label{Gammaintermsofz}
    \Gamma_{+}(t)= \frac{m_0\omega_0(1+z)- m(t)\omega(t)(1-z)}{m_0\omega_0(1+z)+m(t)\omega(t)(1-z)}.
\end{align}

Substituting Eq. \ref{Gammaintermsofz} in \ref{KrylovintermsofGamma}, we get:
\begin{align}
\nonumber
    C_S(t) &=  \frac{1}{4m_0\omega_0 m(t)\omega(t)(1-|z|^2)}\bigg(m_0^2\omega_0^2(1+|z|^2+2 Re(z)) \\ & ~~~~~~ + m(t)^2\omega(t)^2(1+|z|^2-2Re(z))-2m_0\omega_0 m(t)\omega(t)(1-|z|^2)\bigg),\\
    &= \frac{1}{4m(t)\omega(t)}\bigg(\frac{1+|z|^2+2 Re(z)}{1-|z|^2}\bigg)+\frac{1}{4}m(t)\omega(t)\bigg(\frac{1+|z|^2-2Re(z)}{1-|z|^2}\bigg)-\frac{1}{2},
\end{align}
where in the last step we have fixed $m_0=\omega_0=1$.
From Eq. \ref{positionavg} and \ref{momentumavg}, we can express the spread complexity as:
\begin{align}
    C_S(t) &= \frac{1}{2}\bigg(\langle q(t)^2 \rangle+ \langle p(t)^2 \rangle - 1\bigg).
\end{align}

 Using Eq. \ref{circuitcomplexitypandq}, which relates circuit complexity to the expectation values of the position and momentum operators, the following relation between circuit and spread complexity can be established:
 \begin{align}
     C_s(t)= \sinh^2(C(t)).
 \end{align}


Similarly, the rate of change of two different notions of complexity is related as:
\begin{align}
    \frac{dC_S(t)}{dt} &= \frac{1}{2}\bigg(\dot{Q}(t)+\dot{P}(t)\bigg) = \sinh(2C(t))\frac{dC}{dt},
\end{align}
where $\dot{Q}(t)$ and $\dot{P}$ refers to $\frac{d\langle q(t)^2\rangle}{dt}$ and $\frac{d\langle p(t)^2\rangle}{dt}$ respectively.
\\

\textbf{\textit{The growth of spread complexity at early times:}}
It is instructive to study the early time limit of the spread complexity. Using the Krylov basis $\ket{v_n}$, one can expand the state in the Krylov basis as:
\begin{align}
    \ket{\psi(t)}= \sum_{n}\psi_n(t)\ket{v_n},
\end{align}
which produces the following Schrodinger equation:
\begin{align}
\label{schrodingerspreadinitial}
    i\partial_t \psi_n(t) &= a_n(t) \psi_n(t)+b_{n+1}(t)\psi_{n+1}(t)+b_n(t)\psi_{n-1}(t),
\end{align}
where we have the following initial condition:
\begin{align}
    \psi_n(0) &= \delta_{n 0}, ~~~ b_0=0.
\end{align}

By imposing the boundary condition at $t=0$ in Eq. \ref{schrodingerspreadinitial}, we get:
\begin{align}
    i\partial_t \psi_n(0) &= a_0(t) \delta_{n0}+ b_1(t) \delta_{n1} \\
    \dot{\psi}_n(0) &= -i(a_0(t) \delta_{n0}+b_1(t) \delta_{n1})
\end{align}

In writing the above step, $b_0=0$ has been used. 

Taking the derivative of the above equation, we have:
\begin{align}
\nonumber
    \ddot{\psi}_n(0) &= -i(a_n \dot{\psi}_n(0)+b_{n+1}\dot{\psi}_{n+1}(0)+b_n\dot{\psi}_{n-1}(0)) \underbrace{-i (\dot{a}_n(0)\psi_n(0)+\dot{b}_{n+1}(0)\psi_{n+1}(0)+\dot{b}_n(0)\psi_{n-1})}_{\text{Contribution due to the time-dependent Lanczos coefficients }}, \\ \nonumber
    &= -i \bigg(a_n (-i(a_0 \delta_{n0}+b_1 \delta_{n1}))+b_{n+1}(-i(a_0 \delta_{n+1, 0}+b_1 \delta_{n+1, 1})) + b_n (-i(a_0 \delta_{n-1, 0}+b_1 \delta_{n-1, 1}))\bigg) \\ \nonumber& ~~~~~~ -i (\dot{a}_n\delta_{n0}+\dot{b}_{n+1}\delta_{n+1,0}+\dot{b}_n\delta_{n1}),\\ \nonumber
    &= -a_0^2 \delta_{n 0}-a_1b_1\delta_{n1}- a_0 b_0 \delta_{n+1,0}-b_1^2 \delta_{n0}- a_0 b_1 \delta_{n1}-b_1b_2\delta_{n2}-i\dot{a}_0\delta_{n0}-i\dot{b}_1\delta_{n1}, \\ \nonumber
    &= -(a_0^2+b_1^2)\delta_{n0}-(a_1b_1+a_0b_1)\delta_{n1} - b_1b_2 \delta_{n2}-i\dot{a}_0\delta_{n0}-i\dot{b}_1\delta_{n1},\\ 
    &= -(a_0^2+b_1^2+i \dot{a}_0)\delta_{n0}-(a_1b_1+a_0b_1+ i \dot{b}_1)\delta_{n1} - b_1b_2 \delta_{n2}.
\end{align}

From the above equations and the definition of Spread complexity, we have:
\begin{align}
    C(0) &=0, \\
    \dot{C}(0) &=0, \\
    \ddot{C}(0) &= \sum_{n}n \bigg[\ddot{\psi}_n^*\psi_n+\dot{\psi}_n^*\dot{\psi}_n+\dot{\psi}_n^*\dot{\psi}_n+\psi_n^*\ddot{\psi}_n\bigg]=2b_1(t)^2 .
\end{align}
From the above equations, it is understood that only the Lanczos coefficient $b_1$ contributes to the early time growth regime and is independent of the time derivatives of $a_0$ and $b_1$. Moreover, for time-independent case, when the Lanczos coeffcient $b_1$ is constant, spread complexity would show quadratic behaviour in the early time regime, which was already shown in \cite{Huh:2023jxt}. A similar analysis for Krylov operator complexity was done in \cite{Fan:2022xaa}.

However, for time-dependent case, when the Lanczos coefficient is time-dependent, the early time growth of complexity depends on the time dependence of $b_1$.
In terms of $z$, the Lanczos coefficients can be written as:
\begin{align}
\label{anourcase}
    a_n(t) &= \frac{m(t)^2\omega(t)^2+m_0^2\omega_0^2}{2m_0\omega_0m(t)} \bigg(n+\frac{1}{2}\bigg), \\
    \label{bnourcase}
    b_n(t) &= \frac{m(t)^2\omega(t)^2-m_0^2\omega_0^2}{2m_0\omega_0m(t)}(n+1).
\end{align}

In our case, from Eq.\ref{bnourcase}, we have:
\begin{align}
    b_1= 2\frac{m(t)^2\omega(t)^2-m_0^2\omega_0^2}{2m_0\omega_0m(t)}.
\end{align}
Since we fixed $m_0=\omega_0=1$, we have the following:
\begin{align}
    b_1(t)=m(t)\omega(t)^2- \frac{1}{m(t)}.
\end{align}

Therefore, for our consideration, the early time behavior of the spread complexity can be determined from the following equations:
\begin{align}
    C_K(0)=0, ~~~ \dot{C}_K(0)=0, ~~~~ \ddot{C}_K(0)= 2 \bigg(m(t)\omega(t)^2- \frac{1}{m(t)}\bigg)^2.
\end{align}

For explicit $m(t)$ and $\omega(t)$, the above equation can be integrated and the early time behavior of complexity can be exactly determined.

\section{Discussion and Summary}

In this work, we investigated two complementary notions of quantum complexity, viz. circuit complexity and spread complexity for an oscillator with time-dependent mass and frequency. This system serves as the fundamenal building block for understanding field modes in quantum field theory on curved backgrounds, where each mode effectively behaves as an oscillator with time-dependent parameters. Our findings are thus directly relevant to the study of complexity in quantum fields evolving in dynamical spacetimes.

Our established explicit relationships between these measures of quantum complexity and physically interpretable quantities such as particle content, mean energy of the system, and the expectation values of the position and momentum operators. These findings offer a novel way to interpret quantum complexity in terms of physically intuitive quantities, thereby deepening our understanding of information-theoretic aspects of QFT in curved spacetime.

We began our analysis by considering the oscillator to be in the ground state at some initial time $t_0$. The time-evolved state of the oscillator can be completely characterized by a single complex function, the excitation parameter $z(t)$. This parameter measures the deviation of the state from the adiabatic regime and captures the degree of excitations produced by the time dependence of the system. Since both the wave function and all observables can be expressed in terms of $z(t)$, it serves as a complete descriptor of the system's dynamics.

Using this framework, we introduced a well-defined notion of the particle content of the state, the mean value of the energy of the system and expressed the time-dependent expectation values of the position and momentum operators in terms of $z(t)$. Taking the ground state at $t_0$ and the time-evolved state as the reference and target states, respectively, we employed the covariance matrix formalism to compute the circuit complexity. Remarkably, we found that the circuit complexity of the time-evolved state can be expressed entirely in terms of the excitation parameter, revealing a deep connection between circuit complexity and the physical excitation content (particle content) of the system.  

In particular, we find that the circuit complexity of the time-evolved state is directly related to the mean number of quanta and its rate of change with time: 
\begin{align}
    C(t)= \frac{1}{2}{\rm arcosh}\bigg(\mathcal{A}(t) (2\langle n(t) \rangle+1)+ \mathcal{D}(t)\langle \dot{n}(t) \rangle\bigg).
\end{align}
Alternatively, it can also be expressed in terms of the mean value of the energy of the system and its rate of change: 
\begin{align}
    C(t)=\frac{1}{2}{\rm arcosh}\bigg(\mathcal{F}(t)E(t)+ \mathcal{G}(t)\dot{E}(t)\bigg).
\end{align}

Furthermore, we found that the circuit complexity is equivalently related to the combined expectation values of the squared position and momentum operators:
\begin{align}
    C(t)= \frac{1}{2}{\rm arcosh}\bigg( \langle q(t)^2 \rangle+ \langle p(t)^2 \rangle\bigg).
\end{align}

These relations demonstrate that circuit complexity encapsulates detailed information about the dynamical and statistical properties of the system. Given the close correspondence between a time-dependent harmonic oscillator and individual field modes in curved spacetime, our analysis establishes a concrete link between quantum complexity and particle creation phenomenon in QFT on curved backgrounds.

Our analysis of the circuit complexity thus reveals that it can be interpreted directly in terms of physically meaningful quantities. To further strengthen and generalize these connections, it would be valuable to extend the present analysis in several directions. For instance, while we employed the covariance matrix approach in this work, one could alternatively use the wave function approach or the Fubini-Study method introduced in \cite{Chapman:2017rqy}. As demonstrated in \cite{Ali:2018fcz}, different computational schemes for circuit complexity may exhibit varying sensitivities to time-evolution, leading to potential subtle differences in their physical interpretations. Moreover, the ambiguity arising from the choice of the cost function in the definition of circuit complexity can influence the exact relationship between complexity and the physical quantities that reflect the system's dynamics.

Because of all these subtelties, the precise relationship between quantum complexity and the physical quantities such as particle content, mean energy, expectation value of the position and momentum operators remains only partially understood in non-holographic settings. With this motivation, we also investigated the notion of spread complexity, which provides a complementary characterization of the dynamical evolution of quantum states. 

However, the major challenge in studying spread complexity in this context arises from the time dependence of the Hamiltonian. The existing formalism of Krylov or spread complexity has been primarily developed for time-independent Hamiltonians, where the time evolution operator itself generates the Krylov basis. For time-dependent Hamiltonians, however, constructing the Krylov basis and identifying the associated Lanczos coefficients are non-trivial, although some progress has been made in this direction \cite{Takahashi:2024hex}. Nevertheless, when the Hamiltonian is linear in the generators of a Lie algebra, the algebraic structure of the Hamiltonian provides a way forward. The group representation basis in which the Hamiltonian becomes tridiagonal can be identified as the Krylov basis, with time-dependent Lanczos coefficients. 

In particular, for the parametric oscillator whose Hamiltonian can be expressed in terms of the $\mathfrak{su}(1,1)$ generators, the positive discrete series representation basis, in which the Hamiltonian takes a tridiagonal form, naturally plays the role of Krylov basis, thereby enabling a consistent definition of spread complexity for time-dependent Hamiltonians.

Within this framework, we have established a simple yet profound relation between spread complexity and the expectation values of the position and the momentum operator as:
\begin{align}
    C_S(t)= \frac{1}{2}\bigg(\langle q(t)^2 \rangle +\langle p(t)^2\rangle -1\bigg),
\end{align}

 and subsequently uncovered a direct and universal relation between spread and circuit complexity:
\begin{align}
    C_S(t)= \sinh^2(C(t)).
\end{align}

This universal relation constitutes one of the central results of this paper, demonstrating that the circuit complexity of the time-evolved state can be entirely determined by the spread of the state in the Krylov basis. This finding not only bridges two conceptually distinct approaches to quantum complexity but also opens new avenues for exploring the interplay between geometric complexity and Krylov complexity in field theoretic and curved background settings. Our results provide a concrete step towards understanding the emergence and evolution of quantum complexity in dynamical systems and establish it as a fundamental quantity in quantum mechanical and field theoretic systems.

\appendix

\section{A brief review of the Lie algebra decoupling theorem}
\label{appendixA}
The notion of solving the dynamics of a quantum mechanical system is understood as finding a closed-form expression of the time evolution operator and applying it tractably to a quantum state. This can sometimes be a challenging task.
Many mathematical methods have been developed to deal with exponential operators, such as the Magnus expansion \cite{Magnus}, Zassenhaus formula \cite{Zassenhaus}, Suzuki-Trotter decomposition \cite{Suzuki:1976be,suzukimasuo}, among others. However, in order to have full control over the dynamics of the quantum system, it is crucial to go beyond these approximate methods. 

A powerful mathematical technique that can be used to obtain analytical solutions to the dynamics of a quantum system is the \textit{Lie algebra decoupling theorem} developed by Wei and Norman \cite{Weiandnorman}. The formalism was based on the observation that the Lie algebra generated by a Hamiltonian can serve as a convenient basis for studying the system's dynamics. By expressing the time evolution operator in this basis, the problem is reduced to solving a set of scalar differential equations. Whenever these equations admit analytical solutions, the full dynamics of the system can be exactly determined. The Lie algebra decoupling theorem has been widely studied in numerous contexts including open quantum systems. 


The theorem can be interpreted as separating a dynamical problem into directions and the ``strength" (magnitude) of the evolution. The directions correspond to the elements of the Lie algebra, and the coefficients of the elements determine how strongly each algebra element acts on the quantum state. From a more mathematical point of view, the Lie algebra decoupling method effectively transforms the problem of solving an operator-valued linear differential equation into that of solving a coupled system of differential equations. This method is particularly powerful for handling the dynamics of time-dependent Hamiltonians, which are often challenging to treat analytically. However, it is important to note that the Lie algebra decoupling theorem holds only for finite-dimensional Lie algebras. Consequently, this method is best suited for quantum systems whose Hamiltonians are linear on the generators of a representation of some finite dimensional Lie algebra.

For a time-dependent Hamiltonian, the evolution operator satisfies the Schrodinger equation:
\begin{align}
\label{schrodingereqnevolution}
    \frac{dU(t,t_0)}{dt}= -iH(t)U(t,t_0).
\end{align}
The solution to this equation is expressed in the form of a time ordered exponential as:
\begin{align}
    U(t,t_0)= \mathcal{T} \exp\bigg(-i\int_{t_0}^t dt'H(t')\bigg).
\end{align}
The presence of this time ordering makes it difficult to work with this version of the time evolution operator, and the usual approach is iterative, where the time-ordered exponential is expressed in the form of a \textit{Dyson series}. However, for a Hamiltonian that can be written as a sum over $m$ constant operators $K_m$ multiplied by time-dependent coefficients $\lambda_k(t)$:
\begin{align}
    H(t)= \sum_{m}^D\lambda_m(t) K_m, ~~~~~{\rm with}~~~  [K_i,K_j]= \sum_{k} f_{ij}^k K_k,
\end{align}
where $D$ represents the total number of elements of the Lie algebra, the evolution operator can be expressed in a product form. In other words, the existence of a finite dimensional Lie algebra enables the decoupling of the time evolution operator into a product of $n$ operators as:
\begin{align}
    U(t)= U_1(t)U_2(t)....U_n(t),
\end{align}
where each component operator $U_i$ satisfies the equation:
\begin{align}
    \frac{dU_i}{dt}=-i \Gamma_i(t) K_i U_i.
\end{align}
where these $\Gamma_i(t)$'s are time-dependent functions that must be determined.

Interested readers might refer to \cite{Qvarfort:2022yir} for a detailed discussion of the decoupling theorem of the Lie algebra. 


\section{Relation between \texorpdfstring{$\Gamma_3(t)$}{Gamma3(t)} and \texorpdfstring{$\Gamma_{+}(t)$}{GammaPlus(t)}}
\label{appendixB}

In this section, we derive the relation between $\Gamma_3(t)$ and $\Gamma_{+}(t)$ which was necessary to derive the spread complexity. 

The Lie group SU(1,1) is defined as the set of $2\times 2$ matrices $U$ of determinant 1 satisfying $U \epsilon U^{\dagger}=\epsilon$ with:
\begin{align}
    \epsilon= \begin{pmatrix}
        1 & 0\\
        0 & -1
    \end{pmatrix}.
\end{align}

The group elements can be explicitly written as:
\begin{align}
    U= \begin{pmatrix}
       \alpha & \beta \\
       \bar{\beta} & \bar{\alpha}
    \end{pmatrix}
    ~~~~ {\rm with}~~ |\alpha|^2-|\beta|^2=1.
\end{align}

A useful parametrization of the group element is given by:
\begin{align}
\label{genericsu11}
    U(\theta,\phi,\psi)= \begin{pmatrix}
        \cosh(\theta)e^{i\phi} & \sinh(\theta)e^{-i\psi}\\
        \sinh(\theta)e^{i\psi} & \cosh(\theta)e^{-i\phi}
    \end{pmatrix}.
\end{align}

Considering  the following finite dimensional matrix representation for $K_{+}$,$K_{-}$ and $K_3$:
\begin{align}
\label{matrixK}
K_{+}=
    \begin{pmatrix}
 0 & 1 \\
 0 & 0 \\
    \end{pmatrix},
    ~~~~ K_{-}=   \begin{pmatrix}
 0 & 0 \\
 -1 & 0 \\
    \end{pmatrix},
    ~~~~ K_{3} = \frac{1}{2} \begin{pmatrix}
        1 & 0 \\
        0 & -1
    \end{pmatrix}.
\end{align}

The evolution operator in the decomposed form can be written as: 
\begin{align}
    U(t)= e^{\Gamma_{+}(t)K_{+}}e^{\Gamma_{3}(t)K_{3}}e^{\Gamma_{-}(t)K_{-}}.
\end{align}

Using the matrix representation of $K_{\pm}$ and $K_{3}$ written in \ref{matrixK}, we get:
\begin{align}
\label{ourevolution}
    U(t)= e^{-\Gamma_3(t)/2}\begin{pmatrix}
    e^{\Gamma_3(t)}-\Gamma_{-}(t)\Gamma_{+}(t) && \Gamma_{+}(t)\\
    -\Gamma_{-}(t) && 1
    \end{pmatrix}.
\end{align}

Comparing \ref{ourevolution} with \ref{genericsu11}, we get the following relations:
\begin{align}
    |\Gamma_{+}|= |\Gamma_{-}|= \tanh(\theta), ~~~~ \Gamma_{3}= 2 i \phi + 2 \ln\bigg(\frac{1}{\cosh(\theta)}\bigg).
\end{align}

Therefore, we have:
\begin{align}
    Re(\Gamma_3)= -2 \ln(\cosh(\theta))=  \ln(1-|\Gamma_+|^2)
\end{align}

\section{Derivation of the relation between \texorpdfstring{$\Gamma_{+}$}{Gamma+} and the excitation parameter \texorpdfstring{$z$}{z}}
\label{appC}

In this appendix, we provide the derivation of the relation between the time-dependent quantity $\Gamma_{+}$ and the excitation parameter $z$. 

From Eq. \ref{wavefunctionGamma}, we can write the wavefunction of the oscillator as:
\begin{align}
    \psi(q,t)= e^{\frac{\Gamma_3(t)}{4}}\sum_{n=0}^{\infty}\frac{\Gamma_+(t)^n}{2^n n!}\sqrt{2n!}\frac{1}{\sqrt{2^{2n}(2n)!}}\bigg(\frac{m_0\omega_0}{\pi}\bigg)^{1/4}e^{-\frac{m_0\omega_0q^2}{2}}H_{2n}(\sqrt{m_0\omega_0}q).
\end{align}

To relate it to the excitation parameter, we have to suitably express the wavefunction in a form written in \ref{wavefunctionz}. Using the following property of Hermite polynomials \cite{enwiki:1317601709}:
\begin{align}
    H_{2n}(x)= (-4)^n n!L_{n}^{-1/2}(x^2),
\end{align}
where $L_{n}(x)$ is the Laguerre polynomial, the wavefunction $\psi(q,t)$ can be rewritten as:
\begin{align}
    \psi(q,t)= e^{\frac{\Gamma_3(t)}{4}}\bigg(\frac{m_0\omega_0}{\pi}\bigg)^{1/4}e^{-\frac{m_0\omega_0q^2}{2}}\sum_{n=0}^{\infty} (-1)^n \Gamma_{+}(t)^n L_{n}^{-1/2}(m_0\omega_0 q^2) .
\end{align}

Using the generating function for the generalized Laguerre polynomial \cite{enwiki:1310610587}:
\begin{align}
    \sum_{n=0}^{\infty}t^n L_{n}^{(\alpha)}(x)= \frac{1}{(1-t)^{(\alpha+1)}}e^{-t x/(1-t)},
\end{align}

the wavefunction $\psi(q,t)$ can be written as:
\begin{align}
\nonumber
    \psi(q,t) &= e^{\frac{\Gamma_3(t)}{4}}\bigg(\frac{m_0\omega_0}{\pi}\bigg)^{1/4}e^{-\frac{m_0\omega_0q^2}{2}} \frac{1}{\sqrt{1+\Gamma_{+}(t)}} e^{\frac{m_0\omega_0 \Gamma_{+}(t)q^2}{1+\Gamma_{+}(t)}}\\
    &= e^{\frac{\Gamma_3(t)}{4}}\bigg(\frac{m_0\omega_0}{\pi}\bigg)^{1/4}\frac{1}{\sqrt{1+\Gamma_{+}(t)}}\exp\bigg(-\frac{m_0\omega_0}{2}\bigg\{\frac{1-\Gamma_{+}(t)}{1+\Gamma_{+}(t)}\bigg\}q^2\bigg)
\end{align}

Comparing the above equation with Eq. \ref{wavefunctionz}:
\begin{align}
    \psi(q,t)= N(t) \exp\bigg\{-\frac{m(t)\omega(t)}{2}\bigg(\frac{1-z}{1+z}\bigg)q^2\bigg\},
\end{align}

we get:
\begin{align}
    \Gamma_{+}(t)= \frac{m_0\omega_0(1+z)- m(t)\omega(t)(1-z)}{m_0\omega_0(1+z)+m(t)\omega(t)(1-z)}
\end{align}

The normalization factor $N(t)$ can also be identified as:
\begin{align}
    N(t)= e^{\frac{\Gamma_3(t)}{4}}\bigg(\frac{m_0\omega_0}{\pi}\bigg)^{1/4}\frac{1}{\sqrt{1+\Gamma_{+}(t)}}
\end{align}
\section{Application: Oscillator with exponentially varying mass}
\label{sec5}

Let us consider an oscillator with $m=m_0 \exp(2 \gamma t)$ and $\omega(t)= \omega$, This is the \textit{Caldirola Kanai} model \cite{Kanai}. The equation of motion from \ref{equationofmotion} can be written as:
\begin{align}
    \Ddot{q}+2\gamma \dot{q}+\omega^2 q =0,
\end{align}
The above equation can be identified with the equation of motion of a damped harmonic oscillator. 


The quantity $\mu(t)$ that is related to the excitation parameter also satisfies the equation:
\begin{align}
    &\ddot{\mu}+\frac{\dot{m}}{m}\dot{\mu}+\omega^2 \mu=0, \\
    & \ddot{\mu}+2\gamma\dot{\mu}+\omega^2 \mu=0.
\end{align}

Depending on the relative magnitudes of $\gamma$ and $\omega$, three situations arise:
\begin{align}
\mu(t)=
    \begin{cases}
        &  e^{-\gamma t}(A_1 \cos(\Omega t)+ A_2 \sin(\Omega t)), ~~~~~~~~~~~~~ (\omega > \gamma ~~ \text{Underdamped}) \\
        & e^{-\gamma t}(A_1+A_2t), ~~~~~~~~~~~~~~~~~~~~~~~~~~~~~~~~~ (\omega =\gamma ~~ \text{Critically damped}) \\
        & \exp (-\gamma t) (A_1 \exp (\Gamma  t)+A_2 \exp (-\Gamma  t)).~~~~~ (\omega <\gamma ~~ \text{Overdamped})
    \end{cases}
\end{align}

From the relation between $\mu$ and $z$ and using the condition that $z=0$ at $t=0$, since we start from the ground state, the quantity $z$ can be written as:
\begin{align}
    z(t)= \begin{cases}
        \frac{\gamma}{\Omega \cot(\Omega t)+i \omega}, ~~~~~~~~~~~~~ (\omega > \gamma ~~ \text{Underdamped}) \\
        \frac{i\gamma t}{i-\gamma t}, ~~~~~~~~~~~~~~~~~~~~~~ (\omega =\gamma ~~ \text{Critically damped})\\
        \frac{\gamma}{\Gamma \coth(\Gamma t)+i \omega}. ~~~~~~~~~~~~~ (\omega <\gamma ~~ \text{Overdamped})
    \end{cases}
\end{align}

    

With these expressions of the excitation parameter $z$, the mean particle number can be calculated as:
\begin{align}
    \langle n(t) \rangle = \begin{cases}
        \frac{\gamma^2 \sin^2(\Omega t)}{\Omega^2}, ~~~~~~~~~~~~~ (\omega > \gamma ~~ \text{Underdamped}) \\
        \gamma^2 t^2, ~~~~~~~~~~~~~~~~~~~~ (\omega =\gamma ~~ \text{Critically damped}) \\
        \frac{\gamma^2 \sinh^2(\Gamma t)}{\Gamma^2}.~~~~~~~~~~~~~ (\omega <\gamma ~~ \text{Overdamped})
    \end{cases}
\end{align}

\begin{figure}[!ht]
    \centering
    \includegraphics[scale=0.28]{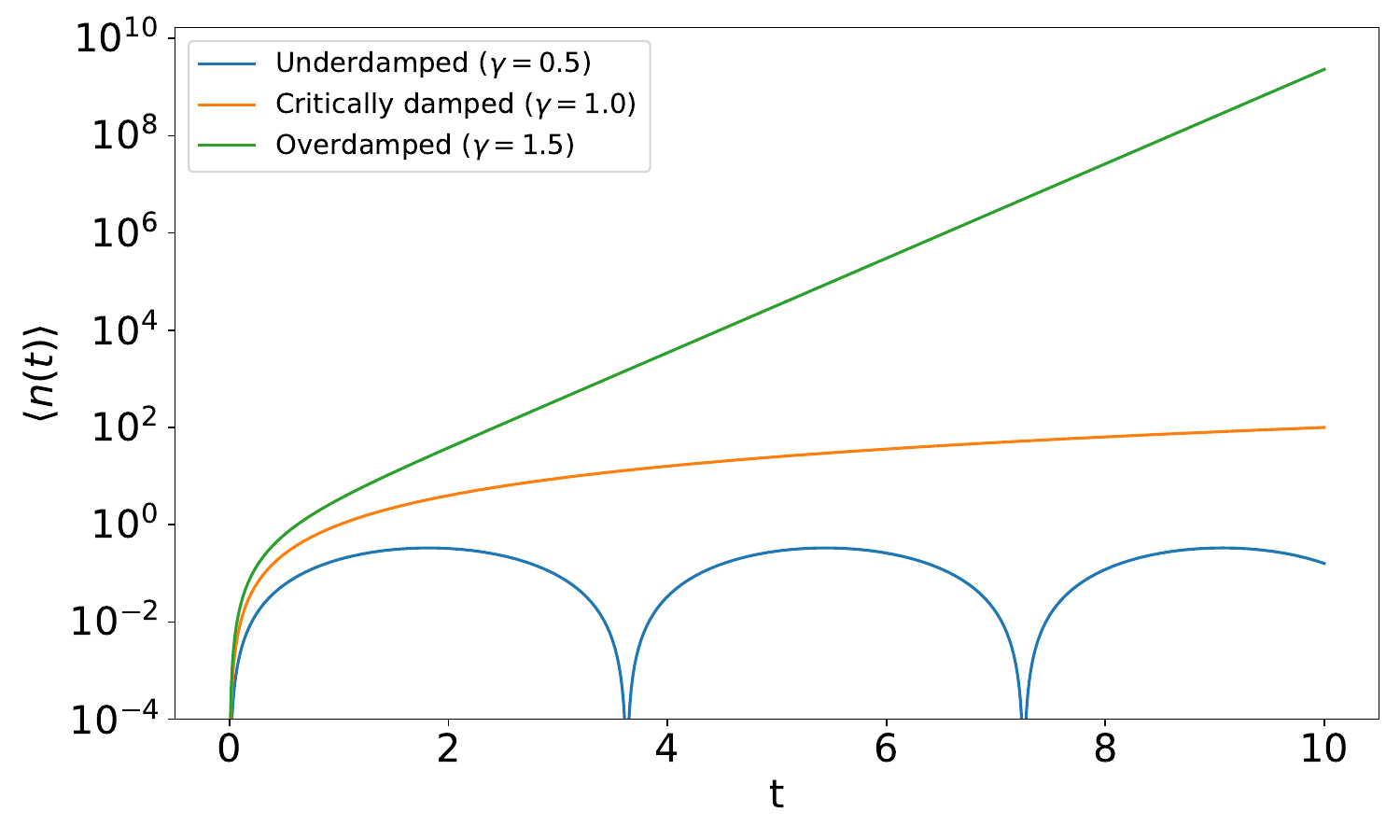}
    \includegraphics[scale=0.28]{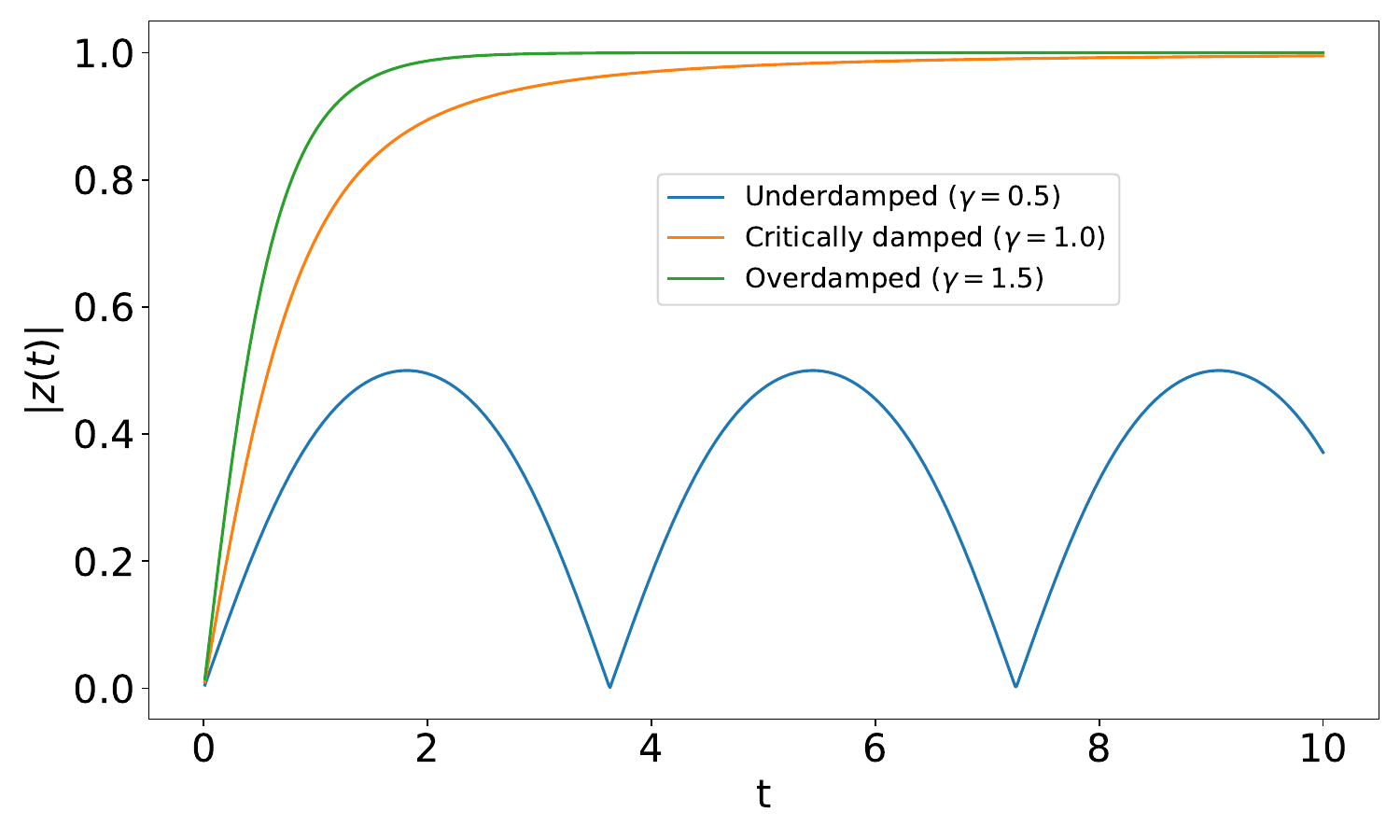}
    \caption{Variation of the mean particle number and the modulus of the excitation parameter with time for different damping scenarios. $\omega$ is fixed to 1.}
    \label{fignvst}
\end{figure}

\begin{figure}[h!]
    \centering
    \includegraphics[scale=0.28]{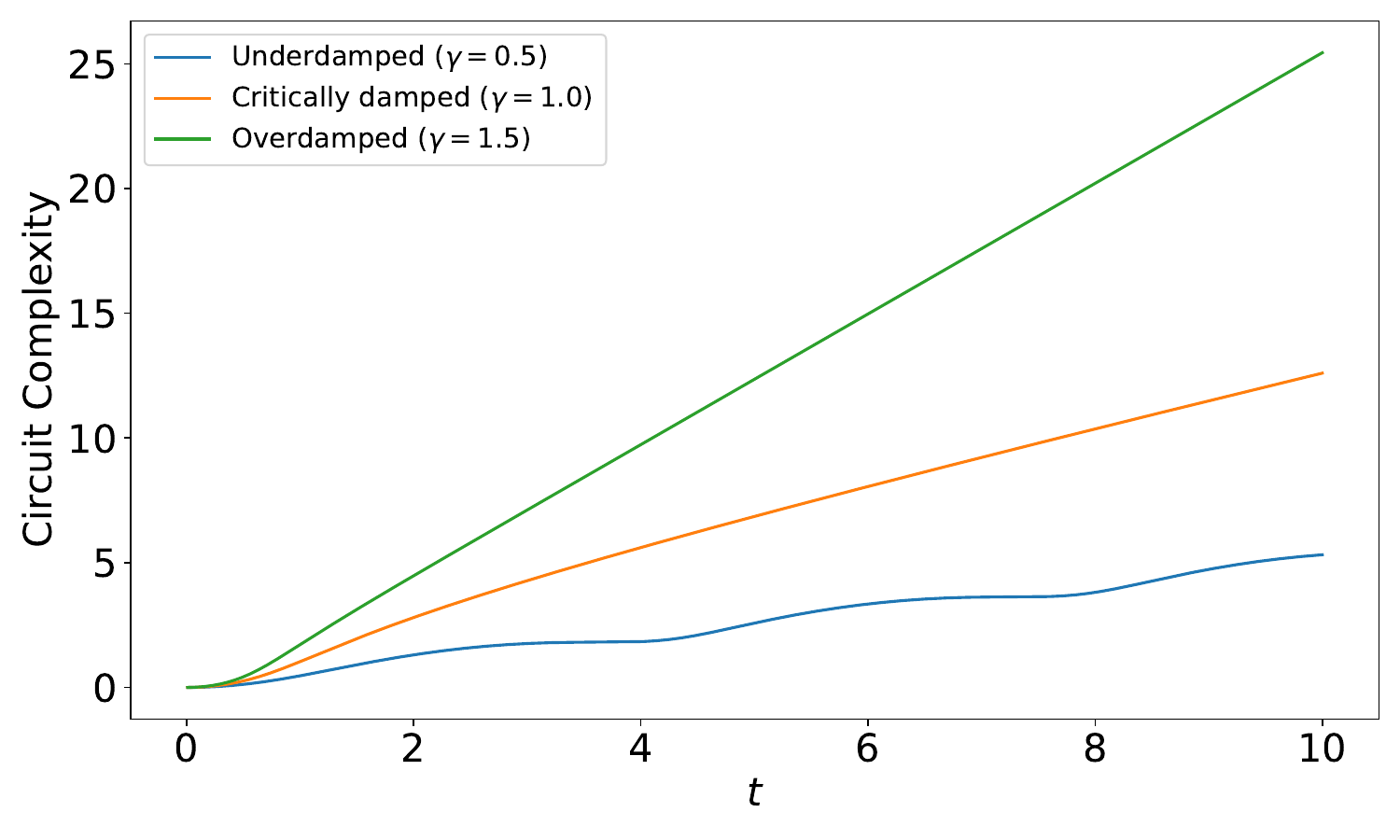}
    \includegraphics[scale=0.28]{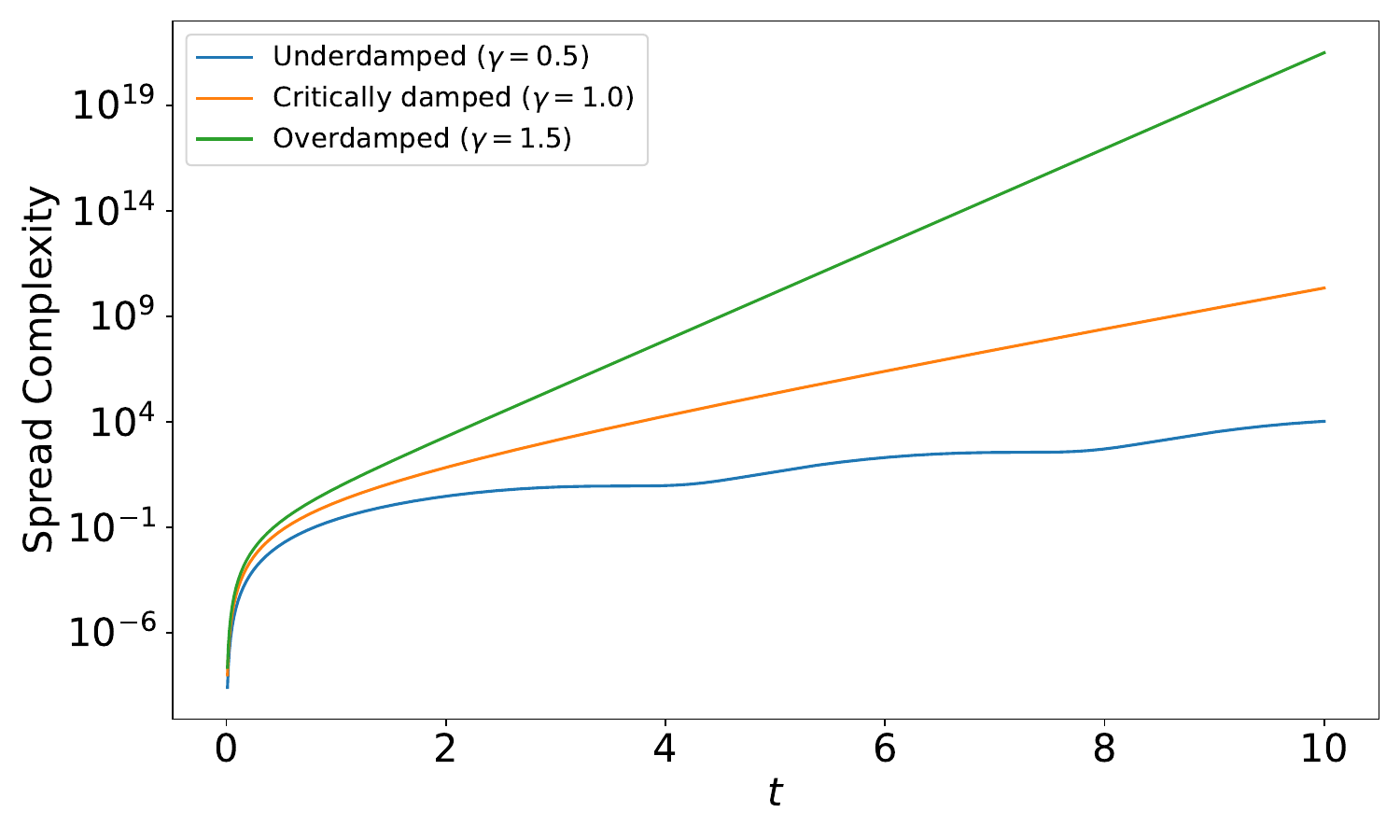}
    \caption{Time evolution of circuit and spread complexity for the different damping scenarios. $\omega$ is fixed to 1.}
    \label{figcircuitcomplexity}
\end{figure}

\begin{figure}[h!]
    \centering
    \includegraphics[scale=0.28]{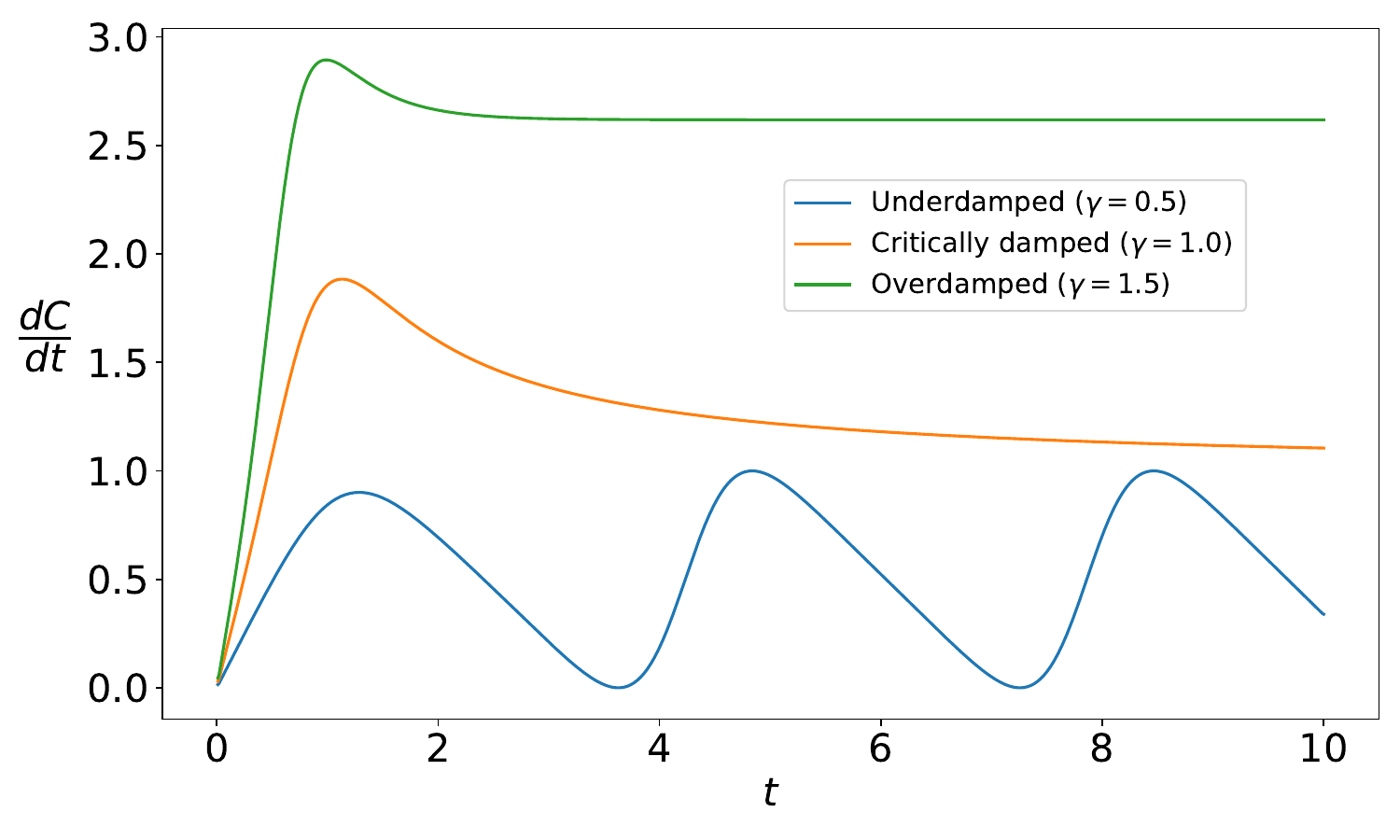}
    \includegraphics[scale=0.28]{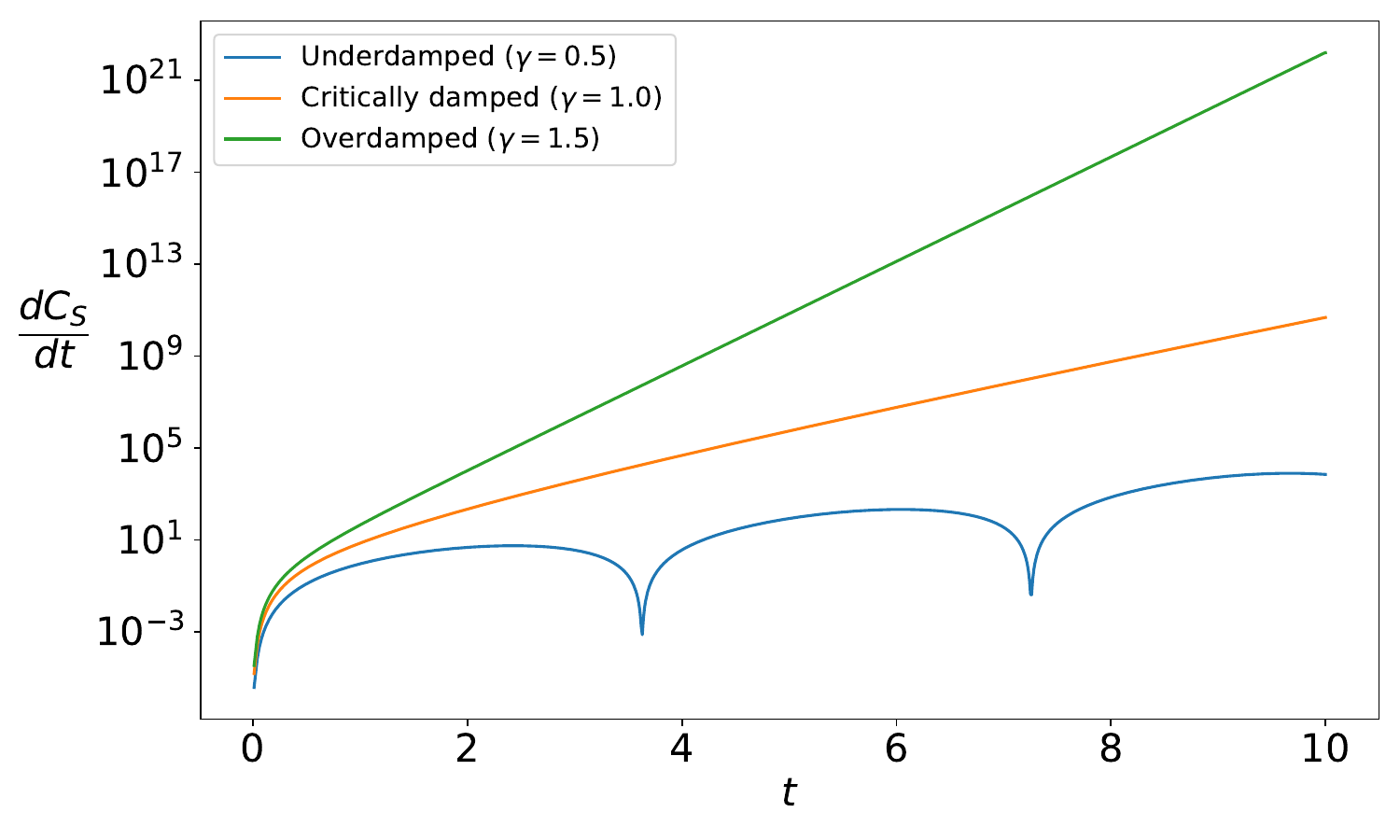}
    \caption{Time evolution of rate of change of circuit and spread complexity for different damping situations. $\omega$ fixed at 1.}
    \label{fig:placeholder}
\end{figure}

\acknowledgments

SC would like to thank the doctoral school of Jagiellonian University for providing fellowship during the work. SC also acknowledges the hospitality of IIT Gandhinagar during his visit, where part of the work was conducted. SC's visit to IIT Gandhinagar was supported by the NAWA PROM programme. SC is grateful to Arpan Bhattacharyya for the discussions, feedback and comments on the draft.





\bibliography{biblio}

\providecommand{\href}[2]{#2}\begingroup\raggedright\begin{thebibliography}{100}

\bibitem{Susskind:2014rva}
L.~Susskind, ``{Computational Complexity and Black Hole Horizons},'' \href{http://dx.doi.org/10.1002/prop.201500092}{{\em Fortsch. Phys.} {\bfseries 64} (2016) 24--43}, \href{http://arxiv.org/abs/1403.5695}{{\ttfamily arXiv:1403.5695 [hep-th]}}. [Addendum: Fortsch.Phys. 64, 44--48 (2016)].

\bibitem{Brown:2015bva}
A.~R. Brown, D.~A. Roberts, L.~Susskind, B.~Swingle, and Y.~Zhao, ``{Holographic Complexity Equals Bulk Action?},'' \href{http://dx.doi.org/10.1103/PhysRevLett.116.191301}{{\em Phys. Rev. Lett.} {\bfseries 116} no.~19, (2016) 191301}, \href{http://arxiv.org/abs/1509.07876}{{\ttfamily arXiv:1509.07876 [hep-th]}}.

\bibitem{Stanford:2014jda}
D.~Stanford and L.~Susskind, ``{Complexity and Shock Wave Geometries},'' \href{http://dx.doi.org/10.1103/PhysRevD.90.126007}{{\em Phys. Rev. D} {\bfseries 90} no.~12, (2014) 126007}, \href{http://arxiv.org/abs/1406.2678}{{\ttfamily arXiv:1406.2678 [hep-th]}}.

\bibitem{Carmi:2016wjl}
D.~Carmi, R.~C. Myers, and P.~Rath, ``{Comments on Holographic Complexity},'' \href{http://dx.doi.org/10.1007/JHEP03(2017)118}{{\em JHEP} {\bfseries 03} (2017) 118}, \href{http://arxiv.org/abs/1612.00433}{{\ttfamily arXiv:1612.00433 [hep-th]}}.

\bibitem{Belin:2021bga}
A.~Belin, R.~C. Myers, S.-M. Ruan, G.~S{\'a}rosi, and A.~J. Speranza, ``{Does Complexity Equal Anything?},'' \href{http://dx.doi.org/10.1103/PhysRevLett.128.081602}{{\em Phys. Rev. Lett.} {\bfseries 128} no.~8, (2022) 081602}, \href{http://arxiv.org/abs/2111.02429}{{\ttfamily arXiv:2111.02429 [hep-th]}}.

\bibitem{Aaronson:2016vto}
S.~Aaronson, ``{The Complexity of Quantum States and Transformations: From Quantum Money to Black Holes},'' in {\em {Proceedings Placeholder}}.
\newblock 7, 2016.
\newblock \href{http://arxiv.org/abs/1607.05256}{{\ttfamily arXiv:1607.05256 [quant-ph]}}.

\bibitem{Jefferson:2017sdb}
R.~Jefferson and R.~C. Myers, ``{Circuit complexity in quantum field theory},'' \href{http://dx.doi.org/10.1007/JHEP10(2017)107}{{\em JHEP} {\bfseries 10} (2017) 107}, \href{http://arxiv.org/abs/1707.08570}{{\ttfamily arXiv:1707.08570 [hep-th]}}.

\bibitem{Caputa:2017yrh}
P.~Caputa, N.~Kundu, M.~Miyaji, T.~Takayanagi, and K.~Watanabe, ``{Liouville Action as Path-Integral Complexity: From Continuous Tensor Networks to AdS/CFT},'' \href{http://dx.doi.org/10.1007/JHEP11(2017)097}{{\em JHEP} {\bfseries 11} (2017) 097}, \href{http://arxiv.org/abs/1706.07056}{{\ttfamily arXiv:1706.07056 [hep-th]}}.

\bibitem{Bhattacharyya:2018wym}
A.~Bhattacharyya, P.~Caputa, S.~R. Das, N.~Kundu, M.~Miyaji, and T.~Takayanagi, ``{Path-Integral Complexity for Perturbed CFTs},'' \href{http://dx.doi.org/10.1007/JHEP07(2018)086}{{\em JHEP} {\bfseries 07} (2018) 086}, \href{http://arxiv.org/abs/1804.01999}{{\ttfamily arXiv:1804.01999 [hep-th]}}.

\bibitem{Caputa:2018kdj}
P.~Caputa and J.~M. Magan, ``{Quantum Computation as Gravity},'' \href{http://dx.doi.org/10.1103/PhysRevLett.122.231302}{{\em Phys. Rev. Lett.} {\bfseries 122} no.~23, (2019) 231302}, \href{http://arxiv.org/abs/1807.04422}{{\ttfamily arXiv:1807.04422 [hep-th]}}.

\bibitem{Bhattacharyya:2018bbv}
A.~Bhattacharyya, A.~Shekar, and A.~Sinha, ``{Circuit complexity in interacting QFTs and RG flows},'' \href{http://dx.doi.org/10.1007/JHEP10(2018)140}{{\em JHEP} {\bfseries 10} (2018) 140}, \href{http://arxiv.org/abs/1808.03105}{{\ttfamily arXiv:1808.03105 [hep-th]}}.

\bibitem{Ali:2018fcz}
T.~Ali, A.~Bhattacharyya, S.~Shajidul~Haque, E.~H. Kim, and N.~Moynihan, ``{Time Evolution of Complexity: A Critique of Three Methods},'' \href{http://dx.doi.org/10.1007/JHEP04(2019)087}{{\em JHEP} {\bfseries 04} (2019) 087}, \href{http://arxiv.org/abs/1810.02734}{{\ttfamily arXiv:1810.02734 [hep-th]}}.

\bibitem{Bhattacharyya:2021cwf}
A.~Bhattacharyya, ``{Circuit complexity and (some of) its applications},'' \href{http://dx.doi.org/10.1142/S0218301321300058}{{\em Int. J. Mod. Phys. E} {\bfseries 30} no.~07, (2021) 2130005}.

\bibitem{Ali:2018aon}
T.~Ali, A.~Bhattacharyya, S.~Shajidul~Haque, E.~H. Kim, and N.~Moynihan, ``{Post-Quench Evolution of Complexity and Entanglement in a Topological System},'' \href{http://dx.doi.org/10.1016/j.physletb.2020.135919}{{\em Phys. Lett. B} {\bfseries 811} (2020) 135919}, \href{http://arxiv.org/abs/1811.05985}{{\ttfamily arXiv:1811.05985 [hep-th]}}.

\bibitem{Osborne:2012mft}
T.~J. Osborne, ``{Hamiltonian complexity},'' \href{http://dx.doi.org/10.1088/0034-4885/75/2/022001}{{\em Rept. Prog. Phys.} {\bfseries 75} no.~2, (2012) 022001}.

\bibitem{Chapman:2017rqy}
S.~Chapman, M.~P. Heller, H.~Marrochio, and F.~Pastawski, ``{Toward a Definition of Complexity for Quantum Field Theory States},'' \href{http://dx.doi.org/10.1103/PhysRevLett.120.121602}{{\em Phys. Rev. Lett.} {\bfseries 120} no.~12, (2018) 121602}, \href{http://arxiv.org/abs/1707.08582}{{\ttfamily arXiv:1707.08582 [hep-th]}}.

\bibitem{Chagnet:2021uvi}
N.~Chagnet, S.~Chapman, J.~de~Boer, and C.~Zukowski, ``{Complexity for Conformal Field Theories in General Dimensions},'' \href{http://dx.doi.org/10.1103/PhysRevLett.128.051601}{{\em Phys. Rev. Lett.} {\bfseries 128} no.~5, (2022) 051601}, \href{http://arxiv.org/abs/2103.06920}{{\ttfamily arXiv:2103.06920 [hep-th]}}.

\bibitem{Chen:2020nlj}
B.~Chen, B.~Czech, and Z.-z. Wang, ``{Query complexity and cutoff dependence of the CFT2 ground state},'' \href{http://dx.doi.org/10.1103/PhysRevD.103.026015}{{\em Phys. Rev. D} {\bfseries 103} no.~2, (2021) 026015}, \href{http://arxiv.org/abs/2004.11377}{{\ttfamily arXiv:2004.11377 [hep-th]}}.

\bibitem{Cotler:2017jue}
J.~Cotler, N.~Hunter-Jones, J.~Liu, and B.~Yoshida, ``{Chaos, Complexity, and Random Matrices},'' \href{http://dx.doi.org/10.1007/JHEP11(2017)048}{{\em JHEP} {\bfseries 11} (2017) 048}, \href{http://arxiv.org/abs/1706.05400}{{\ttfamily arXiv:1706.05400 [hep-th]}}.

\bibitem{Brandao:2019sgy}
F.~G. S.~L. Brand{\~a}o, W.~Chemissany, N.~Hunter-Jones, R.~Kueng, and J.~Preskill, ``{Models of Quantum Complexity Growth},'' \href{http://dx.doi.org/10.1103/PRXQuantum.2.030316}{{\em PRX Quantum} {\bfseries 2} no.~3, (2021) 030316}, \href{http://arxiv.org/abs/1912.04297}{{\ttfamily arXiv:1912.04297 [hep-th]}}.

\bibitem{Haferkamp:2021uxo}
J.~Haferkamp, P.~Faist, N.~B.~T. Kothakonda, J.~Eisert, and N.~Y. Halpern, ``{Linear growth of quantum circuit complexity},'' \href{http://dx.doi.org/10.1038/s41567-022-01539-6}{{\em Nature Phys.} {\bfseries 18} no.~5, (2022) 528--532}, \href{http://arxiv.org/abs/2106.05305}{{\ttfamily arXiv:2106.05305 [quant-ph]}}.

\bibitem{Bouland:2019pvu}
A.~Bouland, B.~Fefferman, and U.~Vazirani, ``{Computational pseudorandomness, the wormhole growth paradox, and constraints on the AdS/CFT duality},'' \href{http://arxiv.org/abs/1910.14646}{{\ttfamily arXiv:1910.14646 [quant-ph]}}.

\bibitem{Munson:2024usy}
A.~Munson, N.~B.~T. Kothakonda, J.~Haferkamp, N.~Y. Halpern, J.~Eisert, and P.~Faist, ``{Complexity-Constrained Quantum Thermodynamics},'' \href{http://dx.doi.org/10.1103/PRXQuantum.6.010346}{{\em PRX Quantum} {\bfseries 6} no.~1, (2025) 010346}, \href{http://arxiv.org/abs/2403.04828}{{\ttfamily arXiv:2403.04828 [quant-ph]}}.

\bibitem{Balasubramanian:2019wgd}
V.~Balasubramanian, M.~Decross, A.~Kar, and O.~Parrikar, ``{Quantum Complexity of Time Evolution with Chaotic Hamiltonians},'' \href{http://dx.doi.org/10.1007/JHEP01(2020)134}{{\em JHEP} {\bfseries 01} (2020) 134}, \href{http://arxiv.org/abs/1905.05765}{{\ttfamily arXiv:1905.05765 [hep-th]}}.

\bibitem{Chowdhury:2023iwg}
S.~Chowdhury, M.~Bojowald, and J.~Mielczarek, ``{Geometric quantum complexity of bosonic oscillator systems},'' \href{http://dx.doi.org/10.1007/JHEP10(2024)048}{{\em JHEP} {\bfseries 10} (2024) 048}, \href{http://arxiv.org/abs/2307.13736}{{\ttfamily arXiv:2307.13736 [quant-ph]}}.

\bibitem{Nielsen_2006}
M.~A. Nielsen, M.~R. Dowling, M.~Gu, and A.~C. Doherty, ``Quantum computation as geometry,'' \href{http://dx.doi.org/10.1126/science.1121541}{{\em Science} {\bfseries 311} no.~5764, (Feb, 2006) 1133--1135}. \url{https://doi.org/10.1126%2Fscience.1121541}.

\bibitem{https://doi.org/10.48550/arxiv.quant-ph/0502070}
M.~A. Nielsen, ``A geometric approach to quantum circuit lower bounds,'' 2005.
\newblock \url{https://arxiv.org/abs/quant-ph/0502070}.

\bibitem{https://doi.org/10.48550/arxiv.quant-ph/0701004}
M.~R. Dowling and M.~A. Nielsen, ``The geometry of quantum computation,'' 2007.
\newblock \url{https://arxiv.org/abs/quant-ph/0701004}.

\bibitem{Parker:2018yvk}
D.~E. Parker, X.~Cao, A.~Avdoshkin, T.~Scaffidi, and E.~Altman, ``{A Universal Operator Growth Hypothesis},'' \href{http://dx.doi.org/10.1103/PhysRevX.9.041017}{{\em Phys. Rev. X} {\bfseries 9} no.~4, (2019) 041017}, \href{http://arxiv.org/abs/1812.08657}{{\ttfamily arXiv:1812.08657 [cond-mat.stat-mech]}}.

\bibitem{Bhattacharyya:2020art}
A.~Bhattacharyya, W.~Chemissany, S.~S. Haque, J.~Murugan, and B.~Yan, ``{The Multi-faceted Inverted Harmonic Oscillator: Chaos and Complexity},'' \href{http://dx.doi.org/10.21468/SciPostPhysCore.4.1.002}{{\em SciPost Phys. Core} {\bfseries 4} (2021) 002}, \href{http://arxiv.org/abs/2007.01232}{{\ttfamily arXiv:2007.01232 [hep-th]}}.

\bibitem{Bhattacharyya:2020kgu}
A.~Bhattacharyya, S.~Das, S.~S. Haque, and B.~Underwood, ``{Rise of cosmological complexity: Saturation of growth and chaos},'' \href{http://dx.doi.org/10.1103/PhysRevResearch.2.033273}{{\em Phys. Rev. Res.} {\bfseries 2} no.~3, (2020) 033273}, \href{http://arxiv.org/abs/2005.10854}{{\ttfamily arXiv:2005.10854 [hep-th]}}.

\bibitem{Bhattacharyya:2020rpy}
A.~Bhattacharyya, S.~Das, S.~Shajidul~Haque, and B.~Underwood, ``{Cosmological Complexity},'' \href{http://dx.doi.org/10.1103/PhysRevD.101.106020}{{\em Phys. Rev. D} {\bfseries 101} no.~10, (2020) 106020}, \href{http://arxiv.org/abs/2001.08664}{{\ttfamily arXiv:2001.08664 [hep-th]}}.

\bibitem{Bhattacharyya:2024duw}
A.~Bhattacharyya, S.~Brahma, S.~S. Haque, J.~S. Lund, and A.~Paul, ``{The early universe as an open quantum system: complexity and decoherence},'' \href{http://dx.doi.org/10.1007/JHEP05(2024)058}{{\em JHEP} {\bfseries 05} (2024) 058}, \href{http://arxiv.org/abs/2401.12134}{{\ttfamily arXiv:2401.12134 [hep-th]}}.

\bibitem{Bhattacharyya:2024rzz}
A.~Bhattacharyya, S.~Brahma, S.~Chowdhury, and X.~Luo, ``{Benchmarking quantum chaos from geometric complexity},'' \href{http://dx.doi.org/10.1007/JHEP03(2025)177}{{\em JHEP} {\bfseries 03} (2025) 177}, \href{http://arxiv.org/abs/2410.18754}{{\ttfamily arXiv:2410.18754 [hep-th]}}.

\bibitem{Bhattacharyya:2025cxv}
A.~Bhattacharyya, S.~Brahma, S.~S. Haque, J.~S. Lund, and A.~Paul, ``{Probing the self-coherence of primordial quantum fluctuations with complexity},'' \href{http://dx.doi.org/10.1088/1475-7516/2025/07/036}{{\em JCAP} {\bfseries 07} (2025) 036}, \href{http://arxiv.org/abs/2502.09739}{{\ttfamily arXiv:2502.09739 [hep-th]}}.

\bibitem{Bhattacharyya:2019kvj}
A.~Bhattacharyya, P.~Nandy, and A.~Sinha, ``{Renormalized Circuit Complexity},'' \href{http://dx.doi.org/10.1103/PhysRevLett.124.101602}{{\em Phys. Rev. Lett.} {\bfseries 124} no.~10, (2020) 101602}, \href{http://arxiv.org/abs/1907.08223}{{\ttfamily arXiv:1907.08223 [hep-th]}}.

\bibitem{Bhattacharyya:2020iic}
A.~Bhattacharyya, S.~S. Haque, and E.~H. Kim, ``{Complexity from the reduced density matrix: a new diagnostic for chaos},'' \href{http://dx.doi.org/10.1007/JHEP10(2021)028}{{\em JHEP} {\bfseries 10} (2021) 028}, \href{http://arxiv.org/abs/2011.04705}{{\ttfamily arXiv:2011.04705 [hep-th]}}.

\bibitem{Bhattacharyya:2022rhm}
A.~Bhattacharyya, T.~Hanif, S.~S. Haque, and A.~Paul, ``{Decoherence, entanglement negativity, and circuit complexity for an open quantum system},'' \href{http://dx.doi.org/10.1103/PhysRevD.107.106007}{{\em Phys. Rev. D} {\bfseries 107} no.~10, (2023) 106007}, \href{http://arxiv.org/abs/2210.09268}{{\ttfamily arXiv:2210.09268 [hep-th]}}.

\bibitem{Bhattacharyya:2021fii}
A.~Bhattacharyya, T.~Hanif, S.~S. Haque, and M.~K. Rahman, ``{Complexity for an open quantum system},'' \href{http://dx.doi.org/10.1103/PhysRevD.105.046011}{{\em Phys. Rev. D} {\bfseries 105} no.~4, (2022) 046011}, \href{http://arxiv.org/abs/2112.03955}{{\ttfamily arXiv:2112.03955 [hep-th]}}.

\bibitem{Erdmenger:2020sup}
J.~Erdmenger, M.~Gerbershagen, and A.-L. Weigel, ``{Complexity measures from geometric actions on Virasoro and Kac-Moody orbits},'' \href{http://dx.doi.org/10.1007/JHEP11(2020)003}{{\em JHEP} {\bfseries 11} (2020) 003}, \href{http://arxiv.org/abs/2004.03619}{{\ttfamily arXiv:2004.03619 [hep-th]}}.

\bibitem{Bhattacharyya:2022ren}
A.~Bhattacharyya, G.~Katoch, and S.~R. Roy, ``{Complexity of warped conformal field theory},'' \href{http://dx.doi.org/10.1140/epjc/s10052-023-11212-8}{{\em Eur. Phys. J. C} {\bfseries 83} no.~1, (2023) 33}, \href{http://arxiv.org/abs/2202.09350}{{\ttfamily arXiv:2202.09350 [hep-th]}}.

\bibitem{Bhattacharyya:2023sjr}
A.~Bhattacharyya and P.~Nandi, ``{Circuit complexity for Carrollian Conformal (BMS) field theories},'' \href{http://dx.doi.org/10.1007/JHEP07(2023)105}{{\em JHEP} {\bfseries 07} (2023) 105}, \href{http://arxiv.org/abs/2301.12845}{{\ttfamily arXiv:2301.12845 [hep-th]}}.

\bibitem{Chowdhury:2024ntx}
S.~Chowdhury, M.~Bojowald, and J.~Mielczarek, ``{Geometric measure of quantum complexity in cosmological systems},'' \href{http://dx.doi.org/10.1103/PhysRevD.111.036036}{{\em Phys. Rev. D} {\bfseries 111} no.~3, (2025) 036036}, \href{http://arxiv.org/abs/2407.01677}{{\ttfamily arXiv:2407.01677 [quant-ph]}}.

\bibitem{Barbon:2019wsy}
J.~L.~F. Barb{\'o}n, E.~Rabinovici, R.~Shir, and R.~Sinha, ``{On The Evolution Of Operator Complexity Beyond Scrambling},'' \href{http://dx.doi.org/10.1007/JHEP10(2019)264}{{\em JHEP} {\bfseries 10} (2019) 264}, \href{http://arxiv.org/abs/1907.05393}{{\ttfamily arXiv:1907.05393 [hep-th]}}.

\bibitem{Rabinovici:2020ryf}
E.~Rabinovici, A.~S{\'a}nchez-Garrido, R.~Shir, and J.~Sonner, ``{Operator complexity: a journey to the edge of Krylov space},'' \href{http://dx.doi.org/10.1007/JHEP06(2021)062}{{\em JHEP} {\bfseries 06} (2021) 062}, \href{http://arxiv.org/abs/2009.01862}{{\ttfamily arXiv:2009.01862 [hep-th]}}.

\bibitem{Nandy:2023brt}
S.~Nandy, B.~Mukherjee, A.~Bhattacharyya, and A.~Banerjee, ``{Quantum state complexity meets many-body scars},'' \href{http://dx.doi.org/10.1088/1361-648X/ad1a7b}{{\em J. Phys. Condens. Matter} {\bfseries 36} no.~15, (2024) 155601}, \href{http://arxiv.org/abs/2305.13322}{{\ttfamily arXiv:2305.13322 [quant-ph]}}.

\bibitem{Dymarsky:2021bjq}
A.~Dymarsky and M.~Smolkin, ``{Krylov complexity in conformal field theory},'' \href{http://dx.doi.org/10.1103/PhysRevD.104.L081702}{{\em Phys. Rev. D} {\bfseries 104} no.~8, (2021) L081702}, \href{http://arxiv.org/abs/2104.09514}{{\ttfamily arXiv:2104.09514 [hep-th]}}.

\bibitem{Avdoshkin:2022xuw}
A.~Avdoshkin, A.~Dymarsky, and M.~Smolkin, ``{Krylov complexity in quantum field theory, and beyond},'' \href{http://dx.doi.org/10.1007/JHEP06(2024)066}{{\em JHEP} {\bfseries 06} (2024) 066}, \href{http://arxiv.org/abs/2212.14429}{{\ttfamily arXiv:2212.14429 [hep-th]}}.

\bibitem{Camargo:2022rnt}
H.~A. Camargo, V.~Jahnke, K.-Y. Kim, and M.~Nishida, ``{Krylov complexity in free and interacting scalar field theories with bounded power spectrum},'' \href{http://dx.doi.org/10.1007/JHEP05(2023)226}{{\em JHEP} {\bfseries 05} (2023) 226}, \href{http://arxiv.org/abs/2212.14702}{{\ttfamily arXiv:2212.14702 [hep-th]}}.

\bibitem{Rabinovici:2022beu}
E.~Rabinovici, A.~S{\'a}nchez-Garrido, R.~Shir, and J.~Sonner, ``{Krylov complexity from integrability to chaos},'' \href{http://dx.doi.org/10.1007/JHEP07(2022)151}{{\em JHEP} {\bfseries 07} (2022) 151}, \href{http://arxiv.org/abs/2207.07701}{{\ttfamily arXiv:2207.07701 [hep-th]}}.

\bibitem{Rabinovici:2021qqt}
E.~Rabinovici, A.~S{\'a}nchez-Garrido, R.~Shir, and J.~Sonner, ``{Krylov localization and suppression of complexity},'' \href{http://dx.doi.org/10.1007/JHEP03(2022)211}{{\em JHEP} {\bfseries 03} (2022) 211}, \href{http://arxiv.org/abs/2112.12128}{{\ttfamily arXiv:2112.12128 [hep-th]}}.

\bibitem{Bhattacharya:2023xjx}
A.~Bhattacharya, P.~P. Nath, and H.~Sahu, ``{Krylov complexity for nonlocal spin chains},'' \href{http://dx.doi.org/10.1103/PhysRevD.109.066010}{{\em Phys. Rev. D} {\bfseries 109} no.~6, (2024) 066010}, \href{http://arxiv.org/abs/2312.11677}{{\ttfamily arXiv:2312.11677 [quant-ph]}}.

\bibitem{Bhattacharjee:2022ave}
B.~Bhattacharjee, P.~Nandy, and T.~Pathak, ``{Krylov complexity in large q and double-scaled SYK model},'' \href{http://dx.doi.org/10.1007/JHEP08(2023)099}{{\em JHEP} {\bfseries 08} (2023) 099}, \href{http://arxiv.org/abs/2210.02474}{{\ttfamily arXiv:2210.02474 [hep-th]}}.

\bibitem{Bhattacharyya:2023dhp}
A.~Bhattacharyya, D.~Ghosh, and P.~Nandi, ``{Operator growth and Krylov complexity in Bose-Hubbard model},'' \href{http://dx.doi.org/10.1007/JHEP12(2023)112}{{\em JHEP} {\bfseries 12} (2023) 112}, \href{http://arxiv.org/abs/2306.05542}{{\ttfamily arXiv:2306.05542 [hep-th]}}.

\bibitem{Balasubramanian:2022tpr}
V.~Balasubramanian, P.~Caputa, J.~M. Magan, and Q.~Wu, ``{Quantum chaos and the complexity of spread of states},'' \href{http://dx.doi.org/10.1103/PhysRevD.106.046007}{{\em Phys. Rev. D} {\bfseries 106} no.~4, (2022) 046007}, \href{http://arxiv.org/abs/2202.06957}{{\ttfamily arXiv:2202.06957 [hep-th]}}.

\bibitem{Bhattacharjee:2022qjw}
B.~Bhattacharjee, S.~Sur, and P.~Nandy, ``{Probing quantum scars and weak ergodicity breaking through quantum complexity},'' \href{http://dx.doi.org/10.1103/PhysRevB.106.205150}{{\em Phys. Rev. B} {\bfseries 106} no.~20, (2022) 205150}, \href{http://arxiv.org/abs/2208.05503}{{\ttfamily arXiv:2208.05503 [quant-ph]}}.

\bibitem{Caputa:2022eye}
P.~Caputa and S.~Liu, ``{Quantum complexity and topological phases of matter},'' \href{http://dx.doi.org/10.1103/PhysRevB.106.195125}{{\em Phys. Rev. B} {\bfseries 106} no.~19, (2022) 195125}, \href{http://arxiv.org/abs/2205.05688}{{\ttfamily arXiv:2205.05688 [hep-th]}}.

\bibitem{Bhattacharya:2022gbz}
A.~Bhattacharya, P.~Nandy, P.~P. Nath, and H.~Sahu, ``{Operator growth and Krylov construction in dissipative open quantum systems},'' \href{http://dx.doi.org/10.1007/JHEP12(2022)081}{{\em JHEP} {\bfseries 12} (2022) 081}, \href{http://arxiv.org/abs/2207.05347}{{\ttfamily arXiv:2207.05347 [quant-ph]}}.

\bibitem{Bhattacharya:2023yec}
A.~Bhattacharya, R.~N. Das, B.~Dey, and J.~Erdmenger, ``{Spread complexity for measurement-induced non-unitary dynamics and Zeno effect},'' \href{http://dx.doi.org/10.1007/JHEP03(2024)179}{{\em JHEP} {\bfseries 03} (2024) 179}, \href{http://arxiv.org/abs/2312.11635}{{\ttfamily arXiv:2312.11635 [hep-th]}}.

\bibitem{Bhattacharya:2023zqt}
A.~Bhattacharya, P.~Nandy, P.~P. Nath, and H.~Sahu, ``{On Krylov complexity in open systems: an approach via bi-Lanczos algorithm},'' \href{http://dx.doi.org/10.1007/JHEP12(2023)066}{{\em JHEP} {\bfseries 12} (2023) 066}, \href{http://arxiv.org/abs/2303.04175}{{\ttfamily arXiv:2303.04175 [quant-ph]}}.

\bibitem{Bhattacharyya:2023grv}
A.~Bhattacharyya, S.~S. Haque, G.~Jafari, J.~Murugan, and D.~Rapotu, ``{Krylov complexity and spectral form factor for noisy random matrix models},'' \href{http://dx.doi.org/10.1007/JHEP10(2023)157}{{\em JHEP} {\bfseries 10} (2023) 157}, \href{http://arxiv.org/abs/2307.15495}{{\ttfamily arXiv:2307.15495 [hep-th]}}.

\bibitem{Carolan:2024wov}
E.~Carolan, A.~Kiely, S.~Campbell, and S.~Deffner, ``{Operator growth and spread complexity in open quantum systems},'' \href{http://dx.doi.org/10.1209/0295-5075/ad5b17}{{\em EPL} {\bfseries 147} no.~3, (2024) 38002}, \href{http://arxiv.org/abs/2404.03529}{{\ttfamily arXiv:2404.03529 [quant-ph]}}.

\bibitem{Liu:2022god}
C.~Liu, H.~Tang, and H.~Zhai, ``{Krylov complexity in open quantum systems},'' \href{http://dx.doi.org/10.1103/PhysRevResearch.5.033085}{{\em Phys. Rev. Res.} {\bfseries 5} no.~3, (2023) 033085}, \href{http://arxiv.org/abs/2207.13603}{{\ttfamily arXiv:2207.13603 [cond-mat.str-el]}}.

\bibitem{Chakrabarti:2025hsb}
N.~Chakrabarti, N.~Nirbhan, and A.~Bhattacharyya, ``{Dynamics of monitored SSH model in Krylov space: from complexity to quantum Fisher information},'' \href{http://dx.doi.org/10.1007/JHEP07(2025)203}{{\em JHEP} {\bfseries 07} (2025) 203}, \href{http://arxiv.org/abs/2502.03434}{{\ttfamily arXiv:2502.03434 [quant-ph]}}.

\bibitem{Bhattacharyya:2025lsc}
A.~Bhattacharyya, S.~Gool, and S.~S. Haque, ``{Krylov Complexity for Open Quantum System: Dissipation and Decoherence},'' \href{http://arxiv.org/abs/2509.14810}{{\ttfamily arXiv:2509.14810 [hep-th]}}.

\bibitem{Baggioli:2025knt}
M.~Baggioli, K.-B. Huh, H.-S. Jeong, X.~Jiang, K.-Y. Kim, and J.~F. Pedraza, ``{Quantum Chaos Diagnostics for Open Quantum Systems from Bi-Lanczos Krylov Dynamics},'' \href{http://arxiv.org/abs/2508.13956}{{\ttfamily arXiv:2508.13956 [hep-th]}}.

\bibitem{Nandy:2024evd}
P.~Nandy, A.~S. Matsoukas-Roubeas, P.~Mart{\'\i}nez-Azcona, A.~Dymarsky, and A.~del Campo, ``{Quantum dynamics in Krylov space: Methods and applications},'' \href{http://dx.doi.org/10.1016/j.physrep.2025.05.001}{{\em Phys. Rept.} {\bfseries 1125-1128} (2025) 1--82}, \href{http://arxiv.org/abs/2405.09628}{{\ttfamily arXiv:2405.09628 [quant-ph]}}.

\bibitem{Baiguera:2025dkc}
S.~Baiguera, V.~Balasubramanian, P.~Caputa, S.~Chapman, J.~Haferkamp, M.~P. Heller, and N.~Y. Halpern, ``{Quantum complexity in gravity, quantum field theory, and quantum information science},'' \href{http://arxiv.org/abs/2503.10753}{{\ttfamily arXiv:2503.10753 [hep-th]}}.

\bibitem{Chapman:2021jbh}
S.~Chapman and G.~Policastro, ``{Quantum computational complexity from quantum information to black holes and back},'' \href{http://dx.doi.org/10.1140/epjc/s10052-022-10037-1}{{\em Eur. Phys. J. C} {\bfseries 82} no.~2, (2022) 128}, \href{http://arxiv.org/abs/2110.14672}{{\ttfamily arXiv:2110.14672 [hep-th]}}.

\bibitem{Susskind:2018tei}
L.~Susskind, ``{Why do Things Fall?},'' \href{http://arxiv.org/abs/1802.01198}{{\ttfamily arXiv:1802.01198 [hep-th]}}.

\bibitem{Susskind:2019ddc}
L.~Susskind, ``{Complexity and Newton's Laws},'' \href{http://dx.doi.org/10.3389/fphy.2020.00262}{{\em Front. in Phys.} {\bfseries 8} (2020) 262}, \href{http://arxiv.org/abs/1904.12819}{{\ttfamily arXiv:1904.12819 [hep-th]}}.

\bibitem{Brown:2018kvn}
A.~R. Brown, H.~Gharibyan, A.~Streicher, L.~Susskind, L.~Thorlacius, and Y.~Zhao, ``{Falling Toward Charged Black Holes},'' \href{http://dx.doi.org/10.1103/PhysRevD.98.126016}{{\em Phys. Rev. D} {\bfseries 98} no.~12, (2018) 126016}, \href{http://arxiv.org/abs/1804.04156}{{\ttfamily arXiv:1804.04156 [hep-th]}}.

\bibitem{Susskind:2020gnl}
L.~Susskind and Y.~Zhao, ``{Complexity and Momentum},'' \href{http://dx.doi.org/10.1007/JHEP03(2021)239}{{\em JHEP} {\bfseries 03} (2021) 239}, \href{http://arxiv.org/abs/2006.03019}{{\ttfamily arXiv:2006.03019 [hep-th]}}.

\bibitem{Magan:2018nmu}
J.~M. Mag{\'a}n, ``{Black holes, complexity and quantum chaos},'' \href{http://dx.doi.org/10.1007/JHEP09(2018)043}{{\em JHEP} {\bfseries 09} (2018) 043}, \href{http://arxiv.org/abs/1805.05839}{{\ttfamily arXiv:1805.05839 [hep-th]}}.

\bibitem{Barbon:2019tuq}
J.~L.~F. Barb{\'o}n, J.~Mart{\'\i}n-Garc{\'\i}a, and M.~Sasieta, ``{Momentum/Complexity Duality and the Black Hole Interior},'' \href{http://dx.doi.org/10.1007/JHEP07(2020)169}{{\em JHEP} {\bfseries 07} (2020) 169}, \href{http://arxiv.org/abs/1912.05996}{{\ttfamily arXiv:1912.05996 [hep-th]}}.

\bibitem{Barbon:2020uux}
J.~L.~F. Barbon, J.~Martin-Garcia, and M.~Sasieta, ``{A Generalized Momentum/Complexity Correspondence},'' \href{http://dx.doi.org/10.1007/JHEP04(2021)250}{{\em JHEP} {\bfseries 04} (2021) 250}, \href{http://arxiv.org/abs/2012.02603}{{\ttfamily arXiv:2012.02603 [hep-th]}}.

\bibitem{Caputa:2024sux}
P.~Caputa, B.~Chen, R.~W. McDonald, J.~Sim{\'o}n, and B.~Strittmatter, ``{Spread Complexity Rate as Proper Momentum},'' \href{http://arxiv.org/abs/2410.23334}{{\ttfamily arXiv:2410.23334 [hep-th]}}.

\bibitem{Fan:2024iop}
Z.-Y. Fan, ``{Momentum-Krylov complexity correspondence},'' \href{http://arxiv.org/abs/2411.04492}{{\ttfamily arXiv:2411.04492 [hep-th]}}.

\bibitem{He:2024pox}
P.-Z. He, ``{Revisit the relationship between spread complexity rate and radial momentum},'' \href{http://arxiv.org/abs/2411.19172}{{\ttfamily arXiv:2411.19172 [hep-th]}}.

\bibitem{Mahajan:2007qc}
G.~Mahajan and T.~Padmanabhan, ``{Particle creation, classicality and related issues in quantum field theory: I. Formalism and toy models},'' \href{http://dx.doi.org/10.1007/s10714-007-0526-z}{{\em Gen. Rel. Grav.} {\bfseries 40} (2008) 661--708}, \href{http://arxiv.org/abs/0708.1233}{{\ttfamily arXiv:0708.1233 [gr-qc]}}.

\bibitem{Mahajan:2007qg}
G.~Mahajan and T.~Padmanabhan, ``{Particle creation, classicality and related issues in quantum field theory: II. Examples from field theory},'' \href{http://dx.doi.org/10.1007/s10714-007-0527-y}{{\em Gen. Rel. Grav.} {\bfseries 40} (2008) 709--747}, \href{http://arxiv.org/abs/0708.1237}{{\ttfamily arXiv:0708.1237 [gr-qc]}}.

\bibitem{Wigner:1932eb}
E.~P. Wigner, ``{On the quantum correction for thermodynamic equilibrium},'' \href{http://dx.doi.org/10.1103/PhysRev.40.749}{{\em Phys. Rev.} {\bfseries 40} (1932) 749--760}.

\bibitem{Albrecht:1992kf}
A.~Albrecht, P.~Ferreira, M.~Joyce, and T.~Prokopec, ``{Inflation and squeezed quantum states},'' \href{http://dx.doi.org/10.1103/PhysRevD.50.4807}{{\em Phys. Rev. D} {\bfseries 50} (1994) 4807--4820}, \href{http://arxiv.org/abs/astro-ph/9303001}{{\ttfamily arXiv:astro-ph/9303001}}.

\bibitem{Matacz:1992mk}
A.~L. Matacz, ``{The Emergence of classical behavior in the quantum fluctuations of a scalar field in an expanding universe},'' \href{http://dx.doi.org/10.1088/0264-9381/10/3/011}{{\em Class. Quant. Grav.} {\bfseries 10} (1993) 509--516}.

\bibitem{Padmanabhan:1986hda}
T.~Padmanabhan and T.~R. Seshadri, ``{Probing the Origin of Large Inhomogeneities in Inflation Using a Toy Quantum Mechanical Model},'' \href{http://dx.doi.org/10.1103/PhysRevD.34.951}{{\em Phys. Rev. D} {\bfseries 34} (1986) 951--958}.

\bibitem{Polarski:1995jg}
D.~Polarski and A.~A. Starobinsky, ``{Semiclassicality and decoherence of cosmological perturbations},'' \href{http://dx.doi.org/10.1088/0264-9381/13/3/006}{{\em Class. Quant. Grav.} {\bfseries 13} (1996) 377--392}, \href{http://arxiv.org/abs/gr-qc/9504030}{{\ttfamily arXiv:gr-qc/9504030}}.

\bibitem{Sriramkumar:2004pj}
L.~Sriramkumar and T.~Padmanabhan, ``{Initial state of matter fields and trans-Planckian physics: Can CMB observations disentangle the two?},'' \href{http://dx.doi.org/10.1103/PhysRevD.71.103512}{{\em Phys. Rev. D} {\bfseries 71} (2005) 103512}, \href{http://arxiv.org/abs/gr-qc/0408034}{{\ttfamily arXiv:gr-qc/0408034}}.

\bibitem{Caves:1985zz}
C.~M. Caves and B.~L. Schumaker, ``{New formalism for two-photon quantum optics. 1. Quadrature phases and squeezed states},'' \href{http://dx.doi.org/10.1103/PhysRevA.31.3068}{{\em Phys. Rev. A} {\bfseries 31} (1985) 3068--3092}.

\bibitem{Schumaker:1986tlu}
B.~L. Schumaker, ``{Quantum mechanical pure states with gaussian wave functions},'' \href{http://dx.doi.org/10.1016/0370-1573(86)90179-1}{{\em Phys. Rept.} {\bfseries 135} no.~6, (1986) 317--408}.

\bibitem{Grishchuk:1989ss}
L.~P. Grishchuk and Y.~V. Sidorov, ``{On the Quantum State of Relic Gravitons},'' \href{http://dx.doi.org/10.1088/0264-9381/6/9/002}{{\em Class. Quant. Grav.} {\bfseries 6} (1989) L161--L165}.

\bibitem{Grishchuk:1990bj}
L.~P. Grishchuk and Y.~V. Sidorov, ``{Squeezed quantum states of relic gravitons and primordial density fluctuations},'' \href{http://dx.doi.org/10.1103/PhysRevD.42.3413}{{\em Phys. Rev. D} {\bfseries 42} (1990) 3413--3421}.

\bibitem{Grishchuk:1994sj}
L.~P. Grishchuk, ``{Density perturbations of quantum mechanical origin and anisotropy of the microwave background},'' \href{http://dx.doi.org/10.1103/PhysRevD.50.7154}{{\em Phys. Rev. D} {\bfseries 50} (1994) 7154--7172}, \href{http://arxiv.org/abs/gr-qc/9405059}{{\ttfamily arXiv:gr-qc/9405059}}.

\bibitem{enwiki:1317601709}
{Wikipedia contributors}, ``Hermite polynomials --- {Wikipedia}{,} the free encyclopedia,'' 2025.
\newblock \url{https://en.wikipedia.org/w/index.php?title=Hermite_polynomials&oldid=1317601709}. [Online; accessed 29-October-2025].

\bibitem{Guo:2018kzl}
M.~Guo, J.~Hernandez, R.~C. Myers, and S.-M. Ruan, ``{Circuit Complexity for Coherent States},'' \href{http://dx.doi.org/10.1007/JHEP10(2018)011}{{\em JHEP} {\bfseries 10} (2018) 011}, \href{http://arxiv.org/abs/1807.07677}{{\ttfamily arXiv:1807.07677 [hep-th]}}.

\bibitem{Hackl:2018ptj}
L.~Hackl and R.~C. Myers, ``{Circuit complexity for free fermions},'' \href{http://dx.doi.org/10.1007/JHEP07(2018)139}{{\em JHEP} {\bfseries 07} (2018) 139}, \href{http://arxiv.org/abs/1803.10638}{{\ttfamily arXiv:1803.10638 [hep-th]}}.

\bibitem{Khan:2018rzm}
R.~Khan, C.~Krishnan, and S.~Sharma, ``{Circuit Complexity in Fermionic Field Theory},'' \href{http://dx.doi.org/10.1103/PhysRevD.98.126001}{{\em Phys. Rev. D} {\bfseries 98} no.~12, (2018) 126001}, \href{http://arxiv.org/abs/1801.07620}{{\ttfamily arXiv:1801.07620 [hep-th]}}.

\bibitem{Balasubramanian:2021mxo}
V.~Balasubramanian, M.~DeCross, A.~Kar, Y.~C. Li, and O.~Parrikar, ``{Complexity growth in integrable and chaotic models},'' \href{http://dx.doi.org/10.1007/JHEP07(2021)011}{{\em JHEP} {\bfseries 07} (2021) 011}, \href{http://arxiv.org/abs/2101.02209}{{\ttfamily arXiv:2101.02209 [hep-th]}}.

\bibitem{Haque:2021hyw}
S.~S. Haque, C.~Jana, and B.~Underwood, ``{Operator complexity for quantum scalar fields and cosmological perturbations},'' \href{http://dx.doi.org/10.1103/PhysRevD.106.063510}{{\em Phys. Rev. D} {\bfseries 106} no.~6, (2022) 063510}, \href{http://arxiv.org/abs/2110.08356}{{\ttfamily arXiv:2110.08356 [hep-th]}}.

\bibitem{Haque:2024ldr}
S.~S. Haque, G.~Jafari, and B.~Underwood, ``{Universal early-time growth in quantum circuit complexity},'' \href{http://dx.doi.org/10.1007/JHEP10(2024)101}{{\em JHEP} {\bfseries 10} (2024) 101}, \href{http://arxiv.org/abs/2406.12990}{{\ttfamily arXiv:2406.12990 [hep-th]}}.

\bibitem{Ali:2019zcj}
T.~Ali, A.~Bhattacharyya, S.~S. Haque, E.~H. Kim, N.~Moynihan, and J.~Murugan, ``{Chaos and Complexity in Quantum Mechanics},'' \href{http://dx.doi.org/10.1103/PhysRevD.101.026021}{{\em Phys. Rev. D} {\bfseries 101} no.~2, (2020) 026021}, \href{http://arxiv.org/abs/1905.13534}{{\ttfamily arXiv:1905.13534 [hep-th]}}.

\bibitem{Lanczos:1950zz}
C.~Lanczos, ``{An iteration method for the solution of the eigenvalue problem of linear differential and integral operators},'' \href{http://dx.doi.org/10.6028/jres.045.026}{{\em J. Res. Natl. Bur. Stand. B} {\bfseries 45} (1950) 255--282}.

\bibitem{Takahashi:2024hex}
K.~Takahashi and A.~del Campo, ``{Krylov Subspace Methods for Quantum Dynamics with Time-Dependent Generators},'' \href{http://dx.doi.org/10.1103/PhysRevLett.134.030401}{{\em Phys. Rev. Lett.} {\bfseries 134} no.~3, (2025) 030401}, \href{http://arxiv.org/abs/2408.08383}{{\ttfamily arXiv:2408.08383 [quant-ph]}}.

\bibitem{Huh:2023jxt}
K.-B. Huh, H.-S. Jeong, and J.~F. Pedraza, ``{Spread complexity in saddle-dominated scrambling},'' \href{http://dx.doi.org/10.1007/JHEP05(2024)137}{{\em JHEP} {\bfseries 05} (2024) 137}, \href{http://arxiv.org/abs/2312.12593}{{\ttfamily arXiv:2312.12593 [hep-th]}}.

\bibitem{Fan:2022xaa}
Z.-Y. Fan, ``{Universal relation for operator complexity},'' \href{http://dx.doi.org/10.1103/PhysRevA.105.062210}{{\em Phys. Rev. A} {\bfseries 105} no.~6, (2022) 062210}, \href{http://arxiv.org/abs/2202.07220}{{\ttfamily arXiv:2202.07220 [quant-ph]}}.

\bibitem{Magnus}
W.~Magnus, ``On the exponential solution of differential equations for a linear operator,'' \href{http://dx.doi.org/https://doi.org/10.1002/cpa.3160070404}{{\em Communications on Pure and Applied Mathematics} {\bfseries 7} no.~4, (1954) 649--673}, \href{http://arxiv.org/abs/https://onlinelibrary.wiley.com/doi/pdf/10.1002/cpa.3160070404}{{\ttfamily https://onlinelibrary.wiley.com/doi/pdf/10.1002/cpa.3160070404}}. \url{https://onlinelibrary.wiley.com/doi/abs/10.1002/cpa.3160070404}.

\bibitem{Zassenhaus}
H.~Zassenhaus, ``Ein verfahren, jeder endlichenp-gruppe einen lie-ring mit der charakteristikp zuzuordnen,'' \href{http://dx.doi.org/10.1007/BF02940757}{{\em Abhandlungen aus dem Mathematischen Seminar der Universität Hamburg} {\bfseries 13} no.~1, (1939) 40--47}, \href{http://arxiv.org/abs/https://doi.org/10.1007/BF02940757}{{\ttfamily https://doi.org/10.1007/BF02940757}}. \url{https://doi.org/10.1007/BF02940757}.

\bibitem{Suzuki:1976be}
M.~Suzuki, ``{Generalized Trotter's Formula and Systematic Approximants of Exponential Operators and Inner Derivations with Applications to Many Body Problems},'' \href{http://dx.doi.org/10.1007/BF01609348}{{\em Commun. Math. Phys.} {\bfseries 51} (1976) 183--190}.

\bibitem{suzukimasuo}
M.~Suzuki, ``Decomposition formulas of exponential operators and lie exponentials with some applications to quantum mechanics and statistical physics,'' \href{http://dx.doi.org/10.1063/1.526596}{{\em Journal of Mathematical Physics} {\bfseries 26} no.~4, (04, 1985) 601--612}, \href{http://arxiv.org/abs/https://pubs.aip.org/aip/jmp/article-pdf/26/4/601/19120226/601\_1\_online.pdf}{{\ttfamily https://pubs.aip.org/aip/jmp/article-pdf/26/4/601/19120226/601\_1\_online.pdf}}. \url{https://doi.org/10.1063/1.526596}.

\bibitem{Weiandnorman}
J.~Wei and E.~Norman, ``Lie algebraic solution of linear differential equations,'' \href{http://dx.doi.org/10.1063/1.1703993}{{\em Journal of Mathematical Physics} {\bfseries 4} no.~4, (04, 1963) 575--581}, \href{http://arxiv.org/abs/https://pubs.aip.org/aip/jmp/article-pdf/4/4/575/19325725/575\_1\_online.pdf}{{\ttfamily https://pubs.aip.org/aip/jmp/article-pdf/4/4/575/19325725/575\_1\_online.pdf}}. \url{https://doi.org/10.1063/1.1703993}.

\bibitem{Qvarfort:2022yir}
S.~Qvarfort and I.~Pikovski, ``{Solving Quantum Dynamics with a Lie-Algebra Decoupling Method},'' \href{http://dx.doi.org/10.1103/PRXQuantum.6.010201}{{\em PRX Quantum} {\bfseries 6} no.~1, (2025) 010201}, \href{http://arxiv.org/abs/2210.11894}{{\ttfamily arXiv:2210.11894 [quant-ph]}}.

\bibitem{enwiki:1310610587}
{Wikipedia contributors}, ``Laguerre polynomials --- {Wikipedia}{,} the free encyclopedia,'' 2025.
\newblock \url{https://en.wikipedia.org/w/index.php?title=Laguerre_polynomials&oldid=1310610587}. [Online; accessed 29-October-2025].

\bibitem{Kanai}
E.~Kanai, ``On the quantization of the dissipative systems*,'' \href{http://dx.doi.org/10.1143/ptp/3.4.440}{{\em Progress of Theoretical Physics} {\bfseries 3} no.~4, (12, 1948) 440--442}, \href{http://arxiv.org/abs/https://academic.oup.com/ptp/article-pdf/3/4/440/5250274/3-4-440.pdf}{{\ttfamily https://academic.oup.com/ptp/article-pdf/3/4/440/5250274/3-4-440.pdf}}. \url{https://doi.org/10.1143/ptp/3.4.440}.

\end{thebibliography}\endgroup
\bibliographystyle{utphys}
\end{document}